\RequirePackage{lineno}
\documentclass[twocolumn,showpacs,preprintnumbers,amsmath,amssymb]{revtex4}

\usepackage{graphicx}
\usepackage{dcolumn}
\usepackage{multirow}
\usepackage{hyperref}  
\usepackage{lineno}
\begin{document}
\def\ra{\ensuremath{\rightarrow}}
\def\la{\ensuremath{\leftarrow}}
\let\rarrow=\ra
\let\larrow=\la
\def\lapprox{\ensuremath{\sim\kern-1em\raise 0.65ex\hbox{$<$}}}
\def\rapprox{\ensuremath{\sim\kern-1em\raise 0.65ex\hbox{$>$}}}
\def\gam{\ensuremath{\gamma}}
\def\rts {\ensuremath{\sqrt{s}}}
\def\stat{\mbox{$\;$(stat.)}}
\def\syst{\mbox{$\;$(syst.)}}
\def\Mtau{\ensuremath{m_{\tau}}}
\def\swsq{\ensuremath{\sin^2\!\theta_{W}}}
\def\swel{\ensuremath{\sin^2\!\theta_{\mathrm{eff}}^{\mathrm{lept}}}}
\def\swsqb{\ensuremath{\sin^2\!\overline{\theta}_{W}}}
\def\swsqon{\ensuremath{\swsq\equiv 1-\mW^2/\mZ^2}}
\def\gv{\ensuremath{g_{\mathrm{V}}}}
\def\ga{\ensuremath{g_{\mathrm{A}}}}
\def\gvbar{\ensuremath{\bar{g}_\mathrm{V}}}
\def\gabar{\ensuremath{\bar{g}_\mathrm{A}}}
\def\antibar#1{\ensuremath{#1\bar{#1}}}
\def\tbar{\ensuremath{\bar{t}}}
\def\ttbar{\antibar{t}}
\def\bbar{\ensuremath{\bar{b}}}
\def\bbbar{\antibar{b}}
\def\cbar{\ensuremath{\bar{c}}}
\def\ccbar{\antibar{c}}
\def\sbar{\ensuremath{\bar{s}}}
\def\ssbar{\antibar{s}}
\def\ubar{\ensuremath{\bar{u}}}
\def\uubar{\antibar{u}}
\def\dbar{\ensuremath{\bar{d}}}
\def\ddbar{\antibar{d}}
\def\fbar{\ensuremath{\bar{f}}}
\def\ffbar{\antibar{f}}
\def\qbar{\ensuremath{\bar{q}}}
\def\qqbar{\antibar{q}}
\def\nbar{\ensuremath{\bar{\nu}}}
\def\nnbar{\antibar{\nu}}
\def\ee{\ensuremath{e^+ e^-}}
\def\epm{\ensuremath{e^{\pm}}}
\def\epem{\ensuremath{e^+ e^-}}
\def\mumu{\ensuremath{\mathrm{\mu^+ \mu^-}}}
\def\tautau{\ensuremath{\mathrm{\tau^+ \tau^-}}}
\let\muchless=\ll
\def\ll{\ensuremath{\ell^+ \ell^-}}
\def\lnu{\ensuremath{\ell \nu}}
\def\Zzero{\ensuremath{Z}}
\def\Zboson{\ensuremath{Z}}
\def\Wplus{\ensuremath{W^+}}
\def\Wminus{\ensuremath{W^-}}
\def\Wboson{\ensuremath{W}}
\def\Wpm{\ensuremath{W^{\pm}}}
\def\Wmp{\ensuremath{W^{\mp}}}
\def\Zzv{\ensuremath{\Zzero^{\textstyle *}}}
\def\Abb{\ensuremath{A_{\bbbar}}}
\def\Acc{\ensuremath{A_{\ccbar}}}
\def\Aqq{\ensuremath{A_{\qqbar}}}
\def\Afb{\ensuremath{A_{{fb}}}}
\def\GZ{\ensuremath{\Gamma_{Z}}}
\def\GW{\ensuremath{\Gamma_{W}}}
\def\GH{\ensuremath{\Gamma_{H}}}
\def\GamHad{\ensuremath{\Gamma_{\mathrm{had}}}}
\def\Gbb{\ensuremath{\Gamma_{\bbbar}}}
\def\Rbb{\ensuremath{R_{\bbbar}}}
\def\Gcc{\ensuremath{\Gamma_{\ccbar}}}
\def\Gvis{\ensuremath{\Gamma_{\mathrm{vis}}}}
\def\Ginv{\ensuremath{\Gamma_{\mathrm{inv}}}}
\def\Bstar{\ensuremath{B^{*}}}
\def\chic{\ensuremath{\raise.4ex\hbox{$\chi$}_{{c}}}}
\def\BoBo{\ensuremath{B^{0}\mbox{--}\bar{B}^{0}}} 
\def\BodBod{\ensuremath{B^{0}_{d}\mbox{--}\bar{B}^{0}_{d}}}
\def\BosBos{\ensuremath{B^{0}_{s}\mbox{--}\bar{B}^{0}_{s}}}
\def\chib{\ensuremath{\raise.4ex\hbox{$\chi$}_{{b}}}}
\def\Epsb  {\ensuremath{\epsilon_{b}}}
\def\Epsc  {\ensuremath{\epsilon_{c}}}
\def\Kstar    {\ensuremath{K^{*}}}
\def\Dstar   {\ensuremath{D^{*}}}
\def\Dsstar   {\ensuremath{D^{**}}}
\def\etpt     {\ensuremath{1/p_{\mathrm{T}} - 1/E_{\mathrm{T}}}} 
\def\etptsig  {\ensuremath{(1/p_{\mathrm{T}} - 1/E_{\mathrm{T}})/(\sigma(1/p_{\mathrm{T}}))}}  
\newcommand{\Bd} {\ensuremath{B_d^0}}
\newcommand{\Bs} {\ensuremath{B_s^0}}
\newcommand{\Bu} {\ensuremath{B_u}}
\newcommand{\Bc} {\ensuremath{B_c}}
\newcommand{\Lb} {\ensuremath{\Lambda_b}}
\newcommand{\btol} {\ensuremath{b \rightarrow \ell}}
\newcommand{\ctol} {\ensuremath{c \rightarrow \ell}}
\newcommand{\btoctol} {\ensuremath{b \rightarrow c \rightarrow \ell}}
\let\psii=\psi
\def\psi{\ensuremath{\psii}}
\def\jpsi{\ensuremath{J/\psi}}
\def\Jpsi{\ensuremath{J/\psi}}
\def\Jee{\ensuremath{\Jpsi\ra\epem}}
\def\Jmm{\ensuremath{\Jpsi\ra\mumu}}
\def\Jmumu{\ensuremath{\Jpsi\ra\mumu}}
\def\Brjl{\ensuremath{\mathrm{Br}(\Jpsi \ra \ll)}}
\def\psip{\ensuremath{\psi^{\sst\prime}}}
\def\alphas{\ensuremath{\alpha_{\mathrm{S}}}}
\def\NF{\ensuremath{N_{\mathrm{F}}}}
\def\NC{\ensuremath{N_{\mathrm{C}}}}
\def\CF{\ensuremath{C_{\mathrm{F}}}}
\def\CA{\ensuremath{C_{\mathrm{A}}}}
\def\TF{\ensuremath{T_{\mathrm{F}}}}
\def\Lms{\ensuremath{\Lambda_{\overline{\mathrm{MS}}}}}
\def\Lmsfive{\ensuremath{\Lambda^{(5)}_{\overline{\mathrm{MS}}}}}
\def\KT{\ensuremath{k_{\perp}}}                                   
\def\Vcb{\ensuremath{\vert V_{cb} \vert}}
\def\Vub{\ensuremath{\vert V_{ub} \vert}}
\def\Vtd{\ensuremath{\vert V_{td} \vert}}
\def\Vts{\ensuremath{\vert V_{ts} \vert}}
\def\Vtb{\ensuremath{\vert V_{tb} \vert}}
\def\Vcs{\ensuremath{\vert V_{cs} \vert}}
\def\Vud{\ensuremath{\vert V_{ud} \vert}}
\def\Vus{\ensuremath{\vert V_{us} \vert}}
\def\Vcd{\ensuremath{\vert V_{cd} \vert}}
\def\LQ{\mathrm{LQ}}
\def\sigmalq{\sigma_{\mathrm{LQ}}}
\def\mlq{M_{\mathrm{LQ}}}
\def\mtlq{M_{\mathrm{LQ}}^{\mathrm{T}}}
\def\LQLQ{\LQ\,\overline{\LQ}}
\def\mmjj{\mu\mu jj}
\def\lljj{lljj}
\def\lnujj{l\nu jj}
\def\mnjj{\mu\nu jj}
\def\eejj{ee jj}
\def\enjj{e\nu jj}
\def\lnjj{l\nu jj}
\def\pt{p_{\mathrm{T}}}
\def\ptall{p_{\mathrm{T}}^{\mathrm{all}}}
\def\pte{p_{\mathrm{T}}^{\mathrm{e}}}
\def\st{S_{\mathrm{T}}}
\def\stl{S_{\mathrm{T}}^{\mathrm{\ell}}}
\def\mt{M_{\mathrm{T}}}
\def\stnu{S_{\mathrm{T}}^{\nu}}
\def\z1myj{p_{\mathrm{T}}^{\mathrm{j}_1}}
\def\ptj{p_{\mathrm{T}}^{\mathrm{j}}}
\def\ptl{p_{\mathrm{T}}^{\mathrm{\ell}}}
\def\phij{\phi^{\mathrm{j}}}
\def\phil{\phi^{\mathrm{\ell}}}
\def\aj1{p_{\mathrm{T}}^{\mathrm{j}_1}}
\def\cj2{p_{\mathrm{T}}^{\mathrm{j}_2}}
\def\dj3{p_{\mathrm{T}}^{\mathrm{j}_3}}
\def\zmyl1pt{p_{\mathrm{T}}^{\mathrm{\ell}_1}}
\def\myl2pt{p_{\mathrm{T}}^{\mathrm{\ell}_2}}
\def\Et{E\mathrm{_T}}
\def\Etcone{E_{\mathrm{T}}^{\mathrm{cone}}}
\def\Ptcone{p_{\mathrm{T}}^{\mathrm{cone}}}
\def\Etcells{\Et^{\mathrm{cells}}}
\def\ptnu{p_{\mathrm{T}}^{\mathrm{\nu}}}
\def\ptm{p_{\mathrm{T}}^{\mathrm{\mu}}}
\def\mll{M_{ll}}
\def\ndsig{N_D^{\mathrm{sig}}}
\def\ndz{N_D^{\mathrm{Z}}}
\def\nmcsig{N_{MC}^{\mathrm{sig}}}
\def\nmcz{N_{MC}^{\mathrm{Z}}}
\def\na{N_{\mathrm{A}}}
\def\nb{N_{\mathrm{B}}}
\def\nc{N_{\mathrm{C}}}
\def\nd{N_{\mathrm{D}}}
\def\pt{\ensuremath{p_{\mathrm{T}}}}
\def\pT{\ensuremath{p_{\mathrm{T}}}}
\def\et{\ensuremath{E_{\mathrm{T}}}}
\def\eT{\ensuremath{E_{\mathrm{T}}}}
\def\ET{\ensuremath{E_{\mathrm{T}}}}
\def\HT{\ensuremath{H_{\mathrm{T}}}}
\def\ptsq{\ensuremath{p^2_{\mathrm{T}}}}
\def\degr{\ensuremath{^\circ}}
\def\abseta{\ensuremath{|\eta|}}
\def\Hgg{\ensuremath{H\to\gamma\gamma}}
\def\mh{\ensuremath{m_h}}
\def\mW{\ensuremath{m_W}}
\def\mZ{\ensuremath{m_Z}}
\def\mH{\ensuremath{m_H}}
\def\mA{\ensuremath{m_A}}
\def\MET{\ensuremath{E_{\mathrm{T}}^{\mathrm{miss}}}}
\def\met{\ensuremath{E_{\mathrm{T}}^{\mathrm{miss}}}}
\def\Wjj{\ensuremath{W \rightarrow jj}}
\def\tjjb{\ensuremath{t \rightarrow jjb}}
\def\Hbb{\ensuremath{H \rightarrow b\bar b}}
\def\Zmm{\ensuremath{Z \rightarrow \mu\mu}}
\def\Zee{\ensuremath{Z \rightarrow ee}}
\def\Zll{\ensuremath{Z \rightarrow \ell\ell}}
\def\Wln{\ensuremath{W \rightarrow \ell\nu}}
\def\Wen{\ensuremath{W \rightarrow e\nu}}
\def\Wmn{\ensuremath{W \rightarrow \mu\nu}}
\def\Hllll{\ensuremath{H \rightarrow ZZ^{(*)} \rightarrow \mu\mu\mu\mu}}
\def\Hmmmm{\ensuremath{H \rightarrow \mu\mu\mu\mu}}
\def\Heeee{\ensuremath{H \rightarrow eeee}}
\def\Amm{\ensuremath{A \rightarrow \mu\mu}}
\def\Ztau{\ensuremath{Z \rightarrow \tau\tau}}
\def\Wtau{\ensuremath{W \rightarrow \tau\nu}}
\def\Atau{\ensuremath{A \rightarrow \tau\tau}}
\def\Htau{\ensuremath{H \rightarrow \tau\tau}}
\def\begL{10$^{31}$~cm$^{-2}$~s$^{-1}$}
\def\lowL{10$^{33}$~cm$^{-2}$~s$^{-1}$}
\def\highL{10$^{34}$~cm$^{-2}$~s$^{-1}$}
\newcommand{\EjetRec}{\ensuremath{E_{\mathrm{rec}}}}
\newcommand{\PjetRec}{\ensuremath{p_{\mathrm{rec}}}}
\newcommand{\EjetTru}{\ensuremath{E_{\mathrm{truth}}}}
\newcommand{\PjetTru}{\ensuremath{p_{\mathrm{truth}}}}
\newcommand{\EjetDM}{\ensuremath{E_{\mathrm{DM}}}}
\newcommand{\Rcone}{\ensuremath{R_{\mathrm{cone}}}}
\def\TeV{\ifmmode {\mathrm{\ Te\kern -0.1em V}}\else
                   \textrm{Te\kern -0.1em V}\fi}
\def\GeV{\ifmmode {\mathrm{\ Ge\kern -0.1em V}}\else
                   \textrm{Ge\kern -0.1em V}\fi}
\def\MeV{\ifmmode {\mathrm{\ Me\kern -0.1em V}}\else
                   \textrm{Me\kern -0.1em V}\fi}
\def\keV{\ifmmode {\mathrm{\ ke\kern -0.1em V}}\else
                   \textrm{ke\kern -0.1em V}\fi}
\def\eV{\ifmmode  {\mathrm{\ e\kern -0.1em V}}\else
                   \textrm{e\kern -0.1em V}\fi}
\let\tev=\TeV
\let\gev=\GeV
\let\mev=\MeV
\let\kev=\keV
\let\ev=\eV

\def\TeVc{\ifmmode {\mathrm{\ Te\kern -0.1em V}/c}\else
                   {\textrm{Te\kern -0.1em V}/$c$}\fi}
\def\GeVc{\ifmmode {\mathrm{\ Ge\kern -0.1em V}/c}\else
                   {\textrm{Ge\kern -0.1em V}/$c$}\fi}
\def\MeVc{\ifmmode {\mathrm{\ Me\kern -0.1em V}/c}\else
                   {\textrm{Me\kern -0.1em V}/$c$}\fi}
\def\keVc{\ifmmode {\mathrm{\ ke\kern -0.1em V}/c}\else
                   {\textrm{ke\kern -0.1em V}/$c$}\fi}
\def\eVc{\ifmmode  {\mathrm{\ e\kern -0.1em V}/c}\else
                   {\textrm{e\kern -0.1em V}/$c$}\fi}
\let\tevc=\TeVc
\let\gevc=\GeVc
\let\mevc=\MeVc
\let\kevc=\keVc
\let\evc=\eVc

\def\TeVcc{\ifmmode {\mathrm{\ Te\kern -0.1em V}/c^2}\else
                   {\textrm{Te\kern -0.1em V}/$c^2$}\fi}
\def\GeVcc{\ifmmode {\mathrm{\ Ge\kern -0.1em V}/c^2}\else
                   {\textrm{Ge\kern -0.1em V}/$c^2$}\fi}
\def\MeVcc{\ifmmode {\mathrm{\ Me\kern -0.1em V}/c^2}\else
                   {\textrm{Me\kern -0.1em V}/$c^2$}\fi}
\def\keVcc{\ifmmode {\mathrm{\ ke\kern -0.1em V}/c^2}\else
                   {\textrm{ke\kern -0.1em V}/$c^2$}\fi}
\def\eVcc{\ifmmode  {\mathrm{\ e\kern -0.1em V}/c^2}\else
                   {\textrm{e\kern -0.1em V}/$c^2$}\fi}
\let\tevcc=\TeVcc
\let\gevcc=\GeVcc
\let\mevcc=\MeVcc
\let\kevcc=\keVcc
\let\evcc=\eVcc

\def\cm{\ifmmode  {\mathrm{\ cm}}\else
                   \textrm{~cm}\fi}

\def\ifb{\mbox{fb$^{-1}$}}.
\def\ipb{\mbox{pb$^{-1}$}}
\def\inb{\mbox{nb$^{-1}$}}
\def\mass#1{\ensuremath{m_{#1#1}}}
\def\twomass#1#2{\ensuremath{m_{#1#2}}} 
\def\Ecm{\ensuremath{E_{\mathrm{cm}}}}
\hyphenation{spectrometer spectro-meter}

\hspace{5.2in} \mbox{Submitted to Phys. Rev. D.}
\begin{flushright}
CERN-PH-EP-2011-050
\end{flushright}

\title{Search for pair production of first or second generation leptoquarks in 
 proton--proton collisions at $\sqrt{s}=7$~TeV using the ATLAS detector at the
 LHC}
\author{The ATLAS Collaboration}
\date{April 22, 2011}

\begin{abstract}
This paper describes searches for the pair production of first or second generation scalar leptoquarks using 35
pb$^{-1}$ of proton-proton collision data recorded by the ATLAS detector at $\sqrt{s}=7$~TeV.  Leptoquarks are seached in events with two oppositely charged muons or electrons and at least two jets, and
in events with one muon or electron, missing transverse momentum and at least two jets.
After event selection, the observed yields are consistent with the predicted backgrounds.
Leptoquark production is excluded at the 95\% CL
 for masses
$\mlq<376 $ (319)$~$GeV and $\mlq<422 $ (362)$~$GeV for first and second generation scalar leptoquarks, respectively,
 when assuming the branching fraction of a leptoquark to a charged lepton is equal to 1.0 (0.5).
\end{abstract}


\maketitle

\section{Introduction}
The Standard Model is extremely successful at describing the 
elementary particles and their interactions, yet it gives no explanation for the
striking symmetry between 
quarks ($q$) and 
leptons ($\ell$). 
This has, in part, been the
motivation for many beyond-the-standard-model theories that posit the existence
of leptoquarks (LQ), particles that carry both lepton and baryon quantum numbers,
which couple to both quarks and leptons, and carry the triplet charge of Quantum Chromodynamics 
(QCD), color.
Theories that predict leptoquarks include models that contain quark
and lepton sub-structure~\cite{theory_substructure2}, theories that seek grand
unification~\cite{theory_gu_all} and models
of extended technicolor~\cite{theory_technicolor}.  Experimental
bounds from searches for flavor-changing-neutral currents  and lepton-family-number
violation place restrictive limits on leptoquark decays to different generations of 
quarks and leptons~\cite{FCNC_results,FCNC_results_2}. 
Direct leptoquark searches at electron--proton colliders have sensitivity to
first generation leptoquarks, but have a non-negligible dependence on the value of the
$LQ-\ell-q$ coupling~\cite{hera}, whereas second and third generation leptoquarks 
are produced via lepton-flavor-violation mechanisms~\cite{LFV}.
The relatively large cross sections~\cite{theory_NLO} for scalar leptoquark production from 
proton-proton collisions lead to the expectation that early LHC data
offer sensitivity to a mass range beyond that probed by other
accelerators. Currently the most stringent limits come from the 
Tevatron~\cite{results_first_gen_limit_d0,results_second_gen_limit_d0}
and from the CMS experiment at the LHC~\cite{results_first_gen_limit_cms,results_second_gen_limit_cms}.

Scalar leptoquarks can be produced in proton--proton collisions either in
leptoquark--antileptoquark pairs ($\LQLQ$) or singly.  
Single LQ production involves an unknown $LQ-\ell-q$ coupling, whereas
the pair production reaction occurs mostly via QCD processes which involve only the
strong coupling constant. Therefore, results from $\LQLQ$ searches
have negligible model dependence, and the only relevant parameter for scalar LQ production is the
leptoquark mass $\mlq$~\cite{footnote1}.
   Final states from  LQ pair production have either two same-flavor oppositely-charged 
leptons and two jets ($lljj$); a lepton,
a neutrino and two jets ($l\nu jj$); or two neutrinos and two jets. 
The leptons $l$ here and throughout the paper are either electrons for 
first generation LQ or muons for the second generation~\cite{footnote3}.
The cross sections for the leptoquark-mediated processes $pp\to\lljj$ and $\lnjj$ 
can be written as $\sigmalq\times\beta^2$ and $\sigmalq\times 2\beta(1-\beta)$, respectively, 
where $\beta$ is the branching fraction for a single leptoquark to decay into a charged lepton and a quark.

\section{Analysis strategy}
This paper reports a search for scalar leptoquark pair production carried out
using 35~pb$^{-1}$ of data recorded by the ATLAS detector during the
2010 LHC proton-proton running period. The analysis is performed 
separately in the  $lljj$ final state  
and in the $l\nu jj$ final state. These searches are combined, leading to final results
presented as a function of $\beta$ and $M_{LQ}$ for the first and second generations. 

Analyses for both final states begin by selecting event samples that have
high acceptance for signal production. At this initial stage, these samples are
dominated by the major backgrounds, $Z$+jets and $\ttbar$
for the $lljj$ case, and $W$+jets and $\ttbar$ for the $l\nu jj$ case. The
samples are then subdivided into orthogonal control and signal regions.

The control regions are
used to validate the background modeling by the Monte Carlo (MC) simulation, and the signal region is
used to search for evidence of LQ production.  The signal region is
defined using an {\it a priori} optimization procedure based on simulated
background and signal events.

A major difference between LQ events and backgrounds is the presence of
jet-lepton (also jet-neutrino in the case of $\lnjj$) pairs coming from the decay
of the parent LQ, giving a peak in the reconstructed jet-lepton mass spectrum for the signal.
The approximate 
reconstruction of these masses
provides the most important variables used to distinguish signal and background
events. In addition, large LQ masses give rise to larger
total measured transverse energy in LQ pair events than is seen for background
events, giving another means to distinguish signal from background. 
Finally, reconstructed boson masses can be used to reject the dominant backgrounds
from $V(V=W,Z)$+jets production where the boson decays into leptons.

In the $lljj$ channel, an average reconstructed leptoquark mass $\overline{M}_{LQ}$ is defined for
each event by computing the average of the masses from the two lepton-jet
combinations in the event. Both possible assignments of the two leptons and the two leading jets to
LQ parents are considered.
The chosen assignment is that which gives the
smallest absolute difference between the two reconstructed masses. The probability
to get the correct pairing with this method is of the order of
90\%. The transverse energy in an event $\stl$ is defined as the scalar sum of the transverse
energy (momentum) of the two electrons (muons) and of the
two leading jets,
$\stl = \zmyl1pt+\myl2pt+\aj1+\cj2$, where the transverse plane is defined as relative
to the beam axis~\cite{footnote2}. The invariant mass of the dilepton pair
$\mll$ provides rejection of the $Z$+jets background.

In the $\lnujj$ channel, LQ mass equivalents are also defined.  The neutrino
transverse momentum $\ptnu$ is inferred from the missing transverse
momentum in the event $\MET$ as described in Section~V.  As in the $lljj$ final state, two pairings
of lepton and jet and $\MET$ and jet are possible.
However, because the component of the neutrino
momentum along the beamline is undetermined, only one mass $\mlq$ from the
charged lepton and a jet can be reconstructed.  The $\MET$ and the remaining jet are
used to compute a transverse mass, $\mtlq =\sqrt{2\ptj\MET(1-\cos\phij)}$
in which $\ptj$ is the transverse momentum of a jet, and $\phij$ is the angle
between the $\ptj$ and the $\MET$ vectors in the transverse plane.
In analogy with the $\lljj$ final state, the chosen pairing is that which gives
the smallest absolute difference between $\mlq$ and $\mtlq$, also resulting
in the correct assignment more than 90\% of the time. The measured
transverse energy is defined as the scalar sum of the lepton transverse momentum,
the missing transverse momentum and the momentum of the two leading jets in the event, 
$\stnu = \zmyl1pt + \MET + \aj1 + \cj2$.  
An additional transverse mass variable 
$\mt = \sqrt{2\zmyl1pt\MET(1-\cos\phil)}$
provides rejection against the dominant
$W$+jets background.  Here, $\ptl$ is the measured lepton transverse momentum,
and $\phil$ is the angle between the $\ptl$ and $\MET$ vectors.

\section{ATLAS detector}
ATLAS~\cite{ATLAS_detPerf, tdr} is a multi-purpose detector with a forward-–backward symmetric cylindrical geometry and nearly 4$\pi$ coverage in solid angle. The three major sub-components of ATLAS are the tracking detector, the calorimeter, and the muon spectrometer. Charged particle tracks and vertices are reconstructed in silicon based tracking detectors that cover $|\eta| <$ 2.5 and transition radiation detectors extending to $|\eta| <$ 2.0. The inner tracking system is immersed in a homogeneous 2~T magnetic field provided by a solenoid. Electron, photon and jet transverse energies are measured in the calorimeter. The ATLAS calorimeter system is segmented into a central barrel and endcaps collectively covering the pseudorapidity range $|\eta| <$ 4.9, and is equipped with both
electromagnetic and hadronic calorimeters. Surrounding the calorimeter, a muon spectrometer with air core toroids, a system of precision tracking chambers, and detectors with triggering capabilities provides precise muon identification and measurements.  A three--level event--triggering system allows for the selection of inclusive muon events with muon transverse momentum ($\ptm$) greater than 13~GeV and inclusive electron events with electron transverse energy ($\Et$) greater than 15~GeV.

\section{Simulated samples}
Event samples processed through a detailed detector
simulation~\cite{atlas-sim} based on GEANT4~\cite{Geant_ref} are used to determine most background and all signal
yields.  
As the recorded data contain
a non-negligible fraction of events with more than one interaction per beam 
crossing, simulated events passing the initial lepton selection criteria 
are reweighted in order to bring the vertex multiplicity distributions 
in data and simulation into agreement. All background samples use ATLAS tune MC09~\cite{mc09_tune}.
  The vector boson processes, $W$+jets and $Z$+jets, are
generated using ALPGEN v2.13~\cite{Alpgen_ref} with CTEQ6L1~\cite{cteq6L1} for the parton distribution functions (PDFs), interfaced to HERWIG v6.510~\cite{herwig_ref} and
JIMMY v4.31~\cite{jimmy_ref}.  Exclusive samples with zero to four additional partons
($n$p with $n=0,\,1,\,2,\,3,\,4$) and an inclusive sample with five or more
additional partons ($n$p with $n=5$) are used.  The cross sections are
computed using the ALPGEN cross sections scaled so that the sum of the $n$p
sample cross sections are equal to the next-to-next-to-leading-order (NNLO) inclusive cross sections
times branching fraction to a single lepton species:
$\sigma(W\to\ell\nu)=10.46 \pm 0.42$~nb and 
$\sigma(Z/\gamma^*\to\ell\ell)=1.070 \pm 0.054$~nb for $M_{ll}>40$~GeV~\cite{b-xsec-Wj-computation,b-xsec-Wj-pub}.
Additional samples for cross checks are generated using SHERPA 1.1.3~\cite{sherpa} with PDF set CTEQ6L1,
and PYTHIA 6.421~\cite{pythia_ref} with PDF MRST 2007 LO*~\cite{mrst07}.
Samples of $\ttbar$ events are produced using MC@NLO v3.41~\cite{mcatnlo_ref} and 
POWHEG 1.01 patch 4~\cite{powheg_ref} with PDF CTEQ6.6M~\cite{cteq66M} interfaced
to HERWIG for parton showering. Cross-sections are scaled to the next-to-next-to-leading-log 
(NNLL) prediction of $165^{+11}_{-16}$~pb~\cite{b-tt-at,b-tt-xsec}. Single top events are
generated using MC@NLO with cross sections of 
3.94~pb, 58.7~pb and 13.1~pb 
for the $s$--, $t$-- and $tW$--channels respectively. Their uncertainties are
 $\pm 10\%$~\cite{singletop1,singletop2}.
Diboson events are generated using HERWIG. Next-to-leading order (NLO) cross
sections are calculated with MCFM~\cite{mcfm}: $44.9\pm2.2$~pb, $18.0\pm1.3$~pb, and
$5.96\pm0.30$~pb for $WW$, $WZ$ ($M_{\ell\ell}>40$~GeV) and $ZZ$
($M_{\ell\ell}>60$~GeV) respectively. Cross section uncertainties take into
account scale and PDF uncertainties, as well as differences
with other generators.

Signal events for LQ masses of 250 to 400~GeV with a 50~GeV binning 
are generated with PYTHIA and tune D6~\cite{d6tune} with cross sections and uncertainties
determined from Ref.~\cite{theory_NLO} using the CTEQ6.6~\cite{cteq_ref} PDF set.
  The unknown $LQ-\ell-q$ coupling value is
set to $0.01\times\sqrt{4\pi\alpha_{EM}}$.  This corresponds to an LQ full width
of less than 1~MeV, and a negligible decay length~\cite{th_hewitt_rev,th_belyaev_rev}.

\section{Object identification}
  Electron candidates are reconstructed from energy clusters
in the electromagnetic calorimeter. Electron identification~\cite{b-medium-ele} is performed using the transverse shower shape,
the longitudinal leakage into the hadronic calorimeter, and the requirement that a good-quality
track
 points to the cluster.
The electron transverse energy is measured in the calorimeter, while
its direction is obtained from the track. 
Further rejection against hadrons is achieved by using the energy deposit patterns in the first
layer of the EM calorimeter. 
In order to suppress the background from photon conversions, a hit 
in the first layer of the pixel detector is required.
Electrons used in this analysis are required to have $\Et>20$~GeV and $|\eta|<2.47$,
with the exclusion of the poorly instrumented region between the barrel and the end-cap
calorimeters at 1.35$<|\eta|<$1.52. A small fraction of events with electrons near problematic regions of the electromagnetic
calorimeter readout is removed.

  Muons selected for this analysis are required to have
$\ptm>20$~GeV and $|\eta|<2.4$. Muon tracks are reconstructed independently 
in the inner detector and in the muon spectrometer, with  a minimum number of hits 
required in each.  A good match is required between the tracks
found in the 
inner tracker and the muon spectrometer.
 In order to reject the cosmic ray background, tracks from muon candidates must extrapolate back to 
the reconstructed event vertex,
satisfying $|d_0|<0.1$~mm and $|z_0|<1$~cm, where $d_0$ is the minimum distance
between the muon trajectory and the event primary vertex in the plane
perpendicular to the beam direction, and $z_0$ is the corresponding distance
parallel to the beam direction.

Finally, both electrons and muons are required to be isolated from other energy
in the calorimeters by imposing $\Etcone/\et<0.2$ and $\Etcone/\ptm<0.25$, 
for electrons and muons, respectively. Here $\Etcone$ is the
transverse energy in the calorimeter in a cone of size $\Delta R \equiv \sqrt{(\Delta \eta)^2 + (\Delta \phi)^2}$ = 0.2 centered on the lepton
direction, excluding the lepton contribution.

The presence of neutrinos is inferred from the missing transverse momentum
$\MET$.  The $\MET$ is defined from the vector sum of the
transverse energy in calorimeter cells included in topological clusters~\cite{ATLAS_detPerf} and the
transverse momentum of the muon, $\MET = -(\Sigma
\vec{\Et}^{\mathrm{cells}} + \vec{p}_{\mathrm{T}}^{\mathrm{\mu}})$.
The clusters are corrected to take into account the different response to
hadrons compared to electrons or photons, as well as dead
material and out-of-cluster energy losses~\cite{met-calib}.

Jets are reconstructed from calorimeter
energy clusters using the anti-$k_t$~\cite{b-antikt} algorithm with a radius
parameter $R=0.4$.  After applying quality requirements based on shower shape and
signal timing with respect to the beam crossing~\cite{JetCleaning}, jets selected for
this analysis must satisfy $\pt>20$~GeV and $|\eta|<2.8$ and must be separated
from leptons by $\Delta R>0.5$.  
Calibrations of lepton and jet transverse momenta, which are mostly derived from control 
samples in data, are applied prior to making the kinematic selections.

\section{Preselection}
Initial selection criteria define event samples with high signal acceptance and yields which are
dominated by the major backgrounds $V$+jets and $\ttbar$. 
Events in both electron and muon
analyses are required to have at least one reconstructed proton--proton
interaction vertex with at least three charged-particle tracks associated to it.
The dilepton channels additionally
require exactly two electrons (muons) with $\et (\ptm)>20$~GeV.  The single lepton channels 
require exactly one electron (muon) with $\ET (\ptm)>20$~GeV, $\MET>25$~GeV and $\mt>40$~GeV.
For the $\enjj$ channel only, a triangle cut $\Delta \phi($jet,$\MET)>$1.5$\times($1-$\MET$/45), where 
$\phi$ is in radians and $\MET$ in GeV is applied, in order to reduce the multijet background contamination.
All analyses require at least two jets satisfying $\pt>20$~GeV. The two highest $\pt$ jets are used to calculate LQ masses.

The acceptance for both signal and background is estimated 
by applying all selection criteria to simulated events.
Expected yields are obtained by scaling the acceptance by 
the predicted cross section and integrated luminosity.

After preselection, good agreement is observed between simulation and data
in both the number of events and the shape of distributions of kinematic variables. 
The determination of the expected backgrounds is discussed in Sections VII and VIII.
The expected and observed yield for the four different channels 
are summarized in Table~\ref{tab:presel}.

\begin{table}
\begin{center}
  \caption{The predicted and observed yields for the preselected sample for all
    channels.
Both statistical and systematic uncertainties are included.
    \label{tab:presel}}
  \begin{tabular}{l||c|c}
   Channel  & Predicted Yield & Observed Yield  \\ \hline
   $\eejj$   & 610$\pm 240$    & 626 \\
   $\enjj$   & 6100$^{+1000}_{-1100}$ & 6088 \\
   $\mmjj$   & 830$^{+200}_{-150}$  & 853 \\
   $\mnjj$   & 9500$\pm2500$ & 9248 \\
  \end{tabular}
\end{center}
\end{table}

\section{Background determination}

Small differences between data and simulation for resolutions and
trigger and reconstruction efficiencies are determined in control data samples
and applied to the simulated events. These corrections influence the obtained
yields by less than 2\%. 
The background arising from lepton misidentification is determined using data-driven
methods in the four channels, as is the contribution from the $Z$+jets background
to $\lljj$ final states. All other backgrounds are modeled with Monte Carlo simulations and 
tested with data in control regions defined
to enhance their contributions, as described in Section~VIII.

Because the $Z$+jets yield in the signal region for the $\lljj$ final state
arises in the tails of distributions, it is determined using a
data-assisted method.  The yield in the signal region $\ndsig$ is calculated as:
\begin{equation}
\ndsig = \frac{\ndz}{\nmcz}\nmcsig,
\end{equation}
\noindent
where $\ndz$ and $\nmcz$ are the observed numbers of events
in data and MC, respectively, in a 20~GeV wide dilepton 
mass window around the nominal $Z$ boson mass, and $\nmcsig$
is the expected $Z$+jets contribution to the signal region (defined
in Table~\ref{t-optvar}).
To estimate a systematic uncertainty, the prediction of $\ndsig$
was derived with and without requirements on the number
of jets in the event, as well as with the different Monte Carlo
generators described in Section~IV. The 3\% background in the 20~GeV mass window from $\ttbar$ and diboson events, estimated with Monte Carlo, is subtracted from the data, as is the $<$1\% contamination from fake leptons. The data-assisted methods (with and without a cut on the number of jets and with various mass window definitions) and the purely Monte Carlo-based estimates agree within 10\%. The largest difference is observed between results obtained with ALPGEN and SHERPA, and is used as a systematic uncertainty. The final estimate is obtained using ALPGEN for $\nmcz$ and $\nmcsig$. 

Jets misidentified as leptons and leptons from 
semileptonic decays of hadrons are referred to as fake leptons. In this paper, events with
fake leptons are referred to as QCD background~\cite{footnote4}.
The QCD background in the $\lljj$ channel 
is estimated
using a fitting method. 
Templates for both real and fake electrons (muons)
are made for the $\Etcone/\et$ ($\Etcone/\pt$) variable, before applying the isolation cut. Simulation is used to
 determine templates for real leptons.
Templates for fake leptons are derived from the data by selecting events 
with exactly one lepton. Events containing $W$ candidates are rejected 
by requiring $\MET<$ 10 GeV. The small residual real lepton contamination is 
estimated from simulated events and is subtracted from the data.
In the $\eejj$ channel, additional templates are made for
the fraction of hits 
associated with transition radiation from electrons.
For both channels, fits with these templates are performed on dilepton
events with two or more jets giving the probability that an event 
contains at least one fake lepton. Fits are made inside the $Z$
 mass window, as well as in the signal region. Both fits indicate a small
expected contribution from fake leptons. In the signal region, an independent
upper limit on the number of fakes is set
by extrapolating from the number of fakes determined in lower jet multiplicity bins and with lower dilepton mass.
The QCD background contribution to the $\mnjj$ analysis is determined
using a scaled control sample method.  In this method, a pair of variables
that cleanly separate real muons from vector boson or leptoquark decay 
from fake muons
is chosen to divide the full sample
into four statistically independent regions.  One of the four regions
corresponds to the signal region, which also
contains muons from vector-boson decays.  The background in the signal region is then
found by scaling the yield in one of the non-signal regions by the 
ratio of yields in the remaining two regions.  This analysis uses $\MET$ and
muon impact parameter $|d_0|$ to define the different regions. The three
background regions are defined as: $|d_0|<0.1$~mm and $\MET<25$~GeV ($A$),
$|d_0|>0.1$~mm and $\MET<25$~GeV ($B$) and $|d_0|>0.1$~mm and $\MET>25$~GeV ($C$). 
The signal region ($D$) is the region with $|d_0|<0.1$~mm and $\MET>25$~GeV.
The correlation between these two variables is less than 10\% in simulated samples, giving stability to the method. 
The yield $\nd$ in the signal region is given by $\nd = \frac{\nc \na}{\nb}$,
where $N$ refer to the observed numbers of events in the four statistically independent regions
after subtraction of the small, residual contributions from events with real muons in the background regions, which are estimated using simulated events.

In the $\enjj$ channel, 
the normalization of the QCD background is
determined from a fit to data of the sum of the $\mt$ distributions for 
QCD events and
all other backgrounds, primarily $W$+jets and $\ttbar$.
This sum is
constrained to equal the total data yield, with the QCD fraction being the
fit parameter. The template for the QCD background shape is determined using a QCD enriched sample
constructed by taking the difference in shapes between two samples, a loose sample selected
using only the electrons passing the trigger requirements, and the nominal sample,
selected using the full electron identification requirements. Residual contamination
from real electrons in the QCD enriched sample is estimated to be 7\%. The latter is estimated with a loose-tight matrix method~\cite{b-tt-at}, and used to
perform a shape--dependent subtraction.
For both muon channels, the background from events containing cosmic ray muons is negligible.

\section{Control regions}

Control samples are used to validate the background determination based on MC.
The two most important control
samples for the $\lljj$ analyses are
(1) $Z$+jet: events for which the dilepton
invariant mass lies within the $Z$ mass window $81\le \mll \le 101$~GeV
and (2) $\ttbar$: events that contain at least two jets and both an electron and a muon selected,
as defined in Section~V. By definition, the $\ttbar$ control region is common for both the $\eejj$ and $\mmjj$ channels.
 The most important
control samples for the $\lnjj$ analyses are 
(1) $W$+2 jets: events with exactly two jets, a charged lepton and $\MET$ such that
 $\mt$ is in the region of the $W$ Jacobian peak (40~GeV $\le$ $\mt$ $\le$ 150~GeV), (2) $W$+3 jets: as in (1) but with at least 3 jets and (3) a $\ttbar$ enriched sample which requires at least 
four jets with $\aj1>50$~GeV, $\cj2>40$~GeV and $\dj3>30$~GeV. 
The expected signal contamination in the control regions is at most 1\%.
The predicted and observed yields in these control samples are shown in
Tables~\ref{t-control-e} and~\ref{t-control-m}. 
Distributions 
 of $\stl$ are shown in
Fig.~\ref{f-eecr} for the $\lljj$ samples in the $Z$+jet and $\ttbar$ control regions, 
and distributions for the reconstructed $\mlq$ are
shown in Fig.~\ref{f-encr} for the $\lnjj$ samples in the $W$+2 jets and the $\ttbar$ control regions. 

\begin{table*}
\begin{center}
\caption{The predicted and observed yields in the control samples for the electron final states. Top refers to both single top and $\ttbar$ events.
Both statistical and systematic uncertainties are included.\label{t-control-e}}
  \begin{tabular}{l||rcl|rcl||rcl|rcl|rcl}
       & \multicolumn{6}{c||}{$\eejj$} & \multicolumn{9}{c}{$\enjj$} \\\hline
   Event          & \multicolumn{6}{c||}{Control Region}& \multicolumn{9}{c}{Control Region}\\
   Source         & \multicolumn{3}{c}{$Z+\ge2$ jets} & \multicolumn{3}{c||}{$\ttbar$}  & \multicolumn{3}{c}{$W+ 2$ jets} & \multicolumn{3}{c}{$W+\ge3$ jet} & \multicolumn{3}{c}{\ttbar} \\ \hline
   $V$+jets       & 150&$\pm$&23 & 0.3&$\pm$&0.1 &2100&$\pm$&$700$ &580&$\pm$&$190$ &$180$&$\pm$&$60$ \\
   Top            & 2.0&$\pm$&0.3    & 24&$\pm$&$4$& 21&$\pm$ &$4$   &44&$\pm$&$9$  &210 &$\pm$&$40$\\
   Diboson        & 2.0&$\pm$&0.3   & 0.8& $\pm$&0.1 & 17&$\pm$&$4$   &8.3 &$\pm$&$1.9$  &2.1&$\pm$&$0.5$ \\
   QCD            & 4.0&$^{+}_{-}$&$^{14.0}_{4.0}$&0.0&$^{+}_{-}$&$^{0.1}_{0.0}$ & 64&$\pm$&14 & 68 &$\pm$&15& 29&$\pm$&7 \\ \hline
   {\bf Total Bkg}& 158&$\pm$&$25$&25&$\pm$&$4$ & 2200&$\pm$&$700$ & 700 &$\pm$&$200$ & 420 &$\pm$&$80$ \\
   Data           & \multicolumn{3}{c|}{140}         & \multicolumn{3}{c||}{22}         & \multicolumn{3}{c|}{2344} & \multicolumn{3}{c|}{722} & \multicolumn{3}{c}{425} \\
\\
  \end{tabular}
\end{center}
\end{table*}

\begin{table*}
\begin{center}
\caption{The predicted and observed yields in the control samples for the muon final states. Top refers to both single top and $\ttbar$ events.
Both statistical and systematic uncertainties are included.\label{t-control-m}}
  \begin{tabular}{l||rcl|rcl||rcl|rcl|rcl}
       & \multicolumn{6}{c||}{$\mmjj$} & \multicolumn{9}{c}{$\mnjj$} \\\hline
   Event          & \multicolumn{6}{c||}{Control Region}& \multicolumn{9}{c}{Control Region}\\
   Source         & \multicolumn{3}{c}{$Z+\ge2$ jets} & \multicolumn{3}{c||}{$\ttbar$}  & \multicolumn{3}{c}{$W+ 2$ jets} & \multicolumn{3}{c}{$W+\ge3$ jet} & \multicolumn{3}{c}{\ttbar} \\ \hline
   $V$+jets       & 190&$\pm$&$24$ &0.3&$\pm$&0.1      &  3300&$\pm$&$1100$ & 900&$\pm$&$300$& 250 &$\pm$&$80$ \\
   Top            & 2.7&$\pm$&$0.5$ &24&$\pm$&$4$    &  14&$\pm$&$3$ & 53&$\pm$&$1$ & 260&$\pm$&$50$ \\
   Diboson        & 0.2&$\pm$&0.1 &0.8 & $\pm$&0.1   &  28&$\pm$&$6$ & 14&$\pm$&$3$ & 3.0&$\pm$&$0.7$ \\
   QCD            & 6.0&$^{+}_{-}$&$^{11.0}_{6.0}$& 0.0&$^{+}_{-}$&$^{0.1}_{0.0}$&300 &$\pm$& 100& 130 &$\pm$&50 &  54 &$\pm$&32 \\ \hline
   {\bf Total Bkg}& 200&$\pm$&$25$ & 25&$\pm$&$4$ &  3600&$\pm$&$1100$ & 1100 &$\pm$&$330$ & 570 &$\pm$&$120$ \\
   Data           & \multicolumn{3}{c|}{216}         & \multicolumn{3}{c||}{22}         & \multicolumn{3}{c|}{3588} & \multicolumn{3}{c|}{1120} & \multicolumn{3}{c}{547} \\
  \end{tabular}
\end{center}
\end{table*}

\begin{figure}[ht!]
  \includegraphics[width=3.5truein]{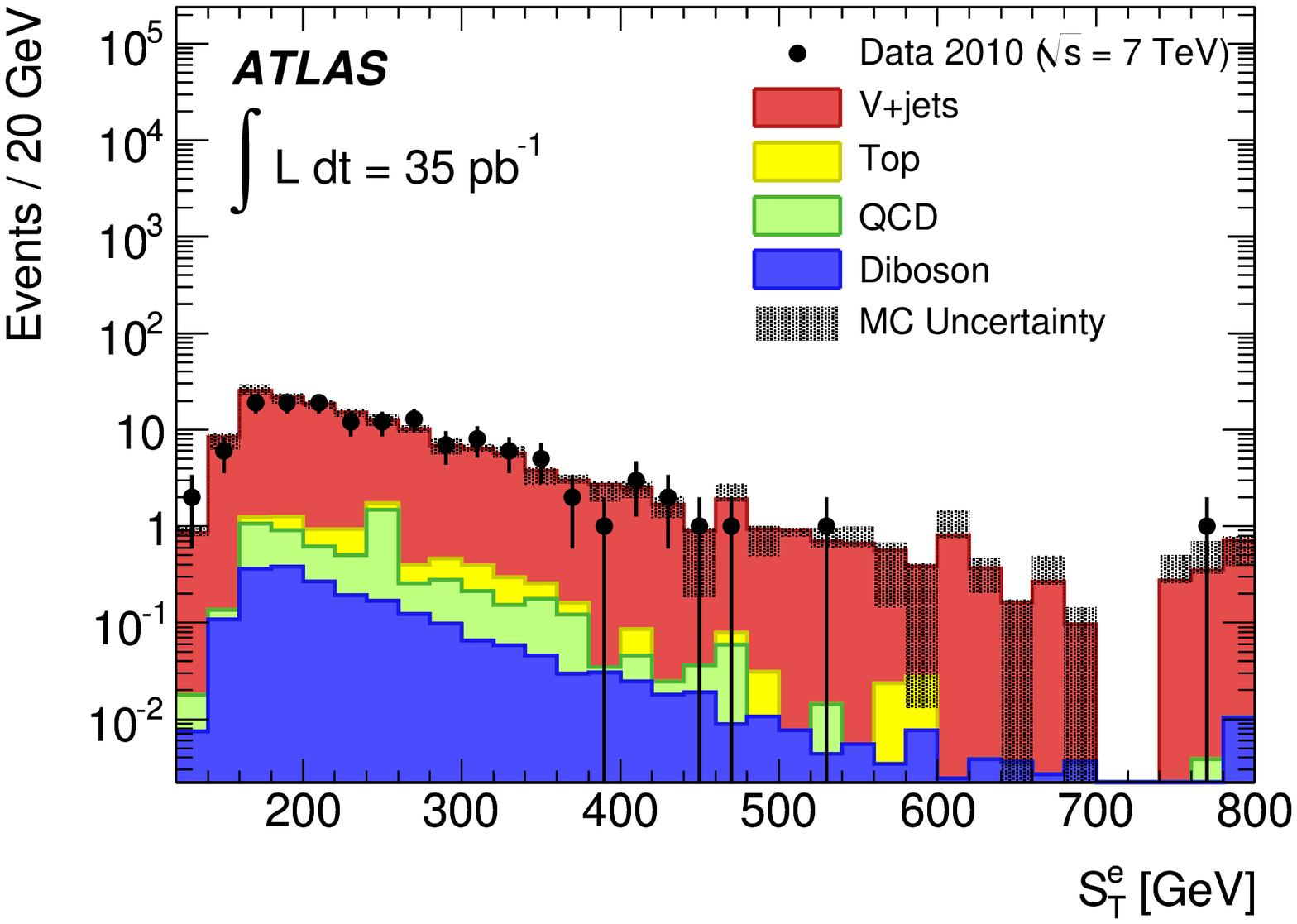}
\includegraphics[width=3.5truein]{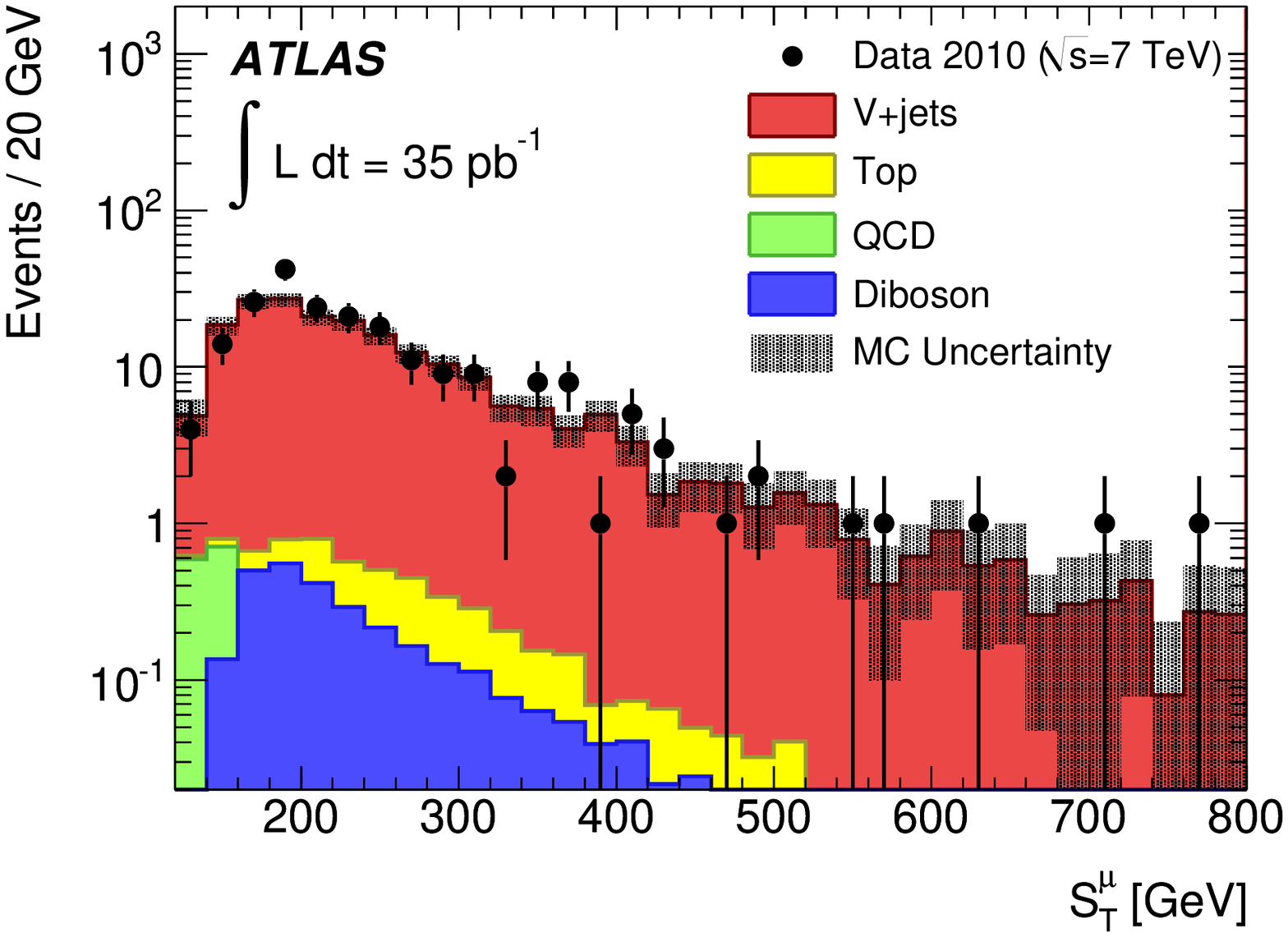}
\includegraphics[width=3.5truein]{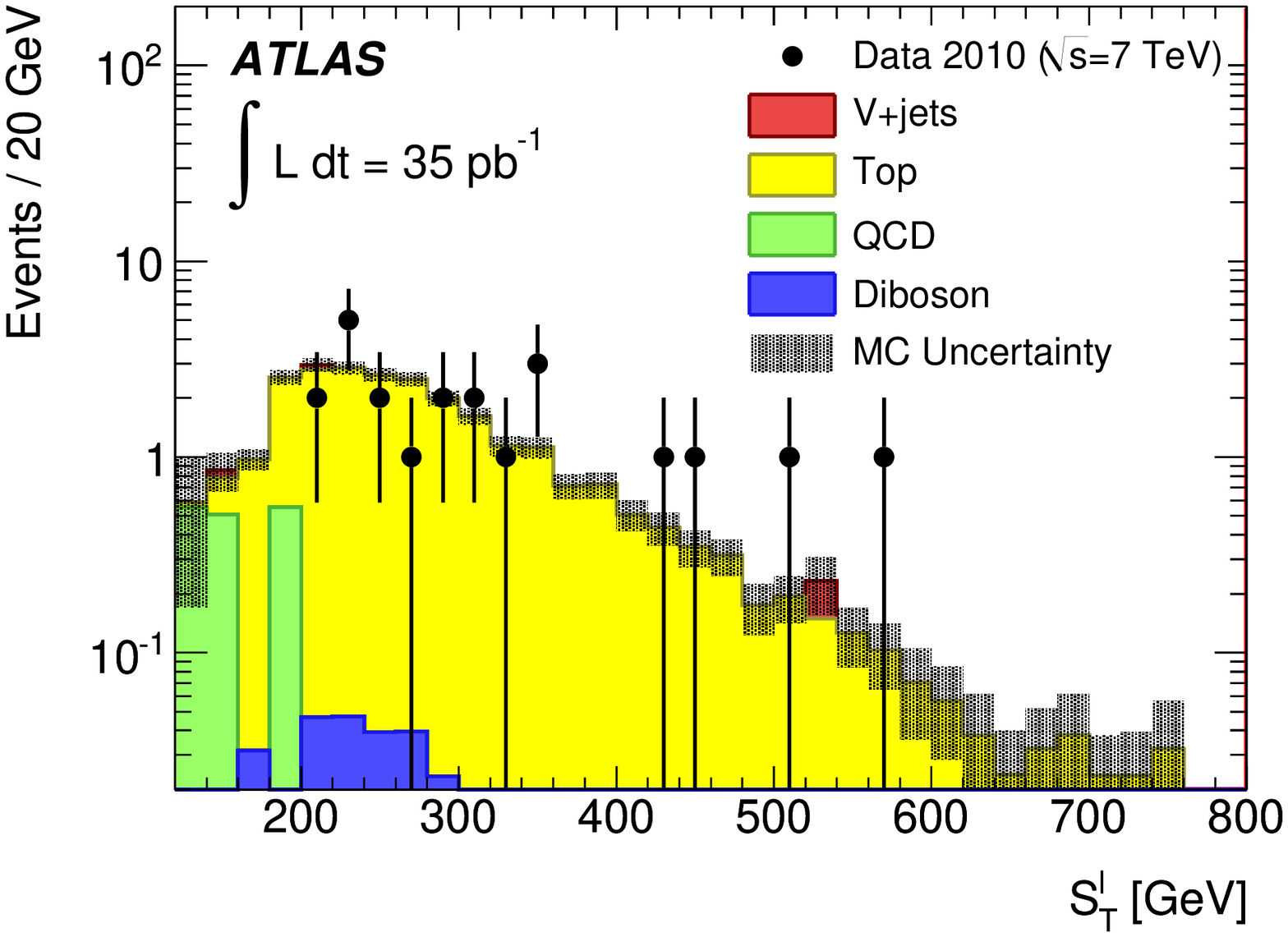}
  \caption{$\stl$ distributions for the
   $\eejj$ (top) and the $\mmjj$ final states (middle) in the Z control region. 
The bottom figure shows the $\stl$ distribution for the dilepton $\ttbar$ control region.
The data are indicated by the points
and the Standard Model backgrounds are shown with cumulative distributions.
The QCD background is estimated from data, while the other background contributions
are obtained from simulated samples. The top background includes both $\ttbar$ and single top events.
 The expected contribution from a potential LQ signal is negligible
and is not included in the figure. The MC uncertainty band shows the combined statistical 
and systematic uncertainties on the prediction in each bin.
   \label{f-eecr}}
\end{figure}

\begin{figure*}[ht!]
  \includegraphics[width=3.5truein]{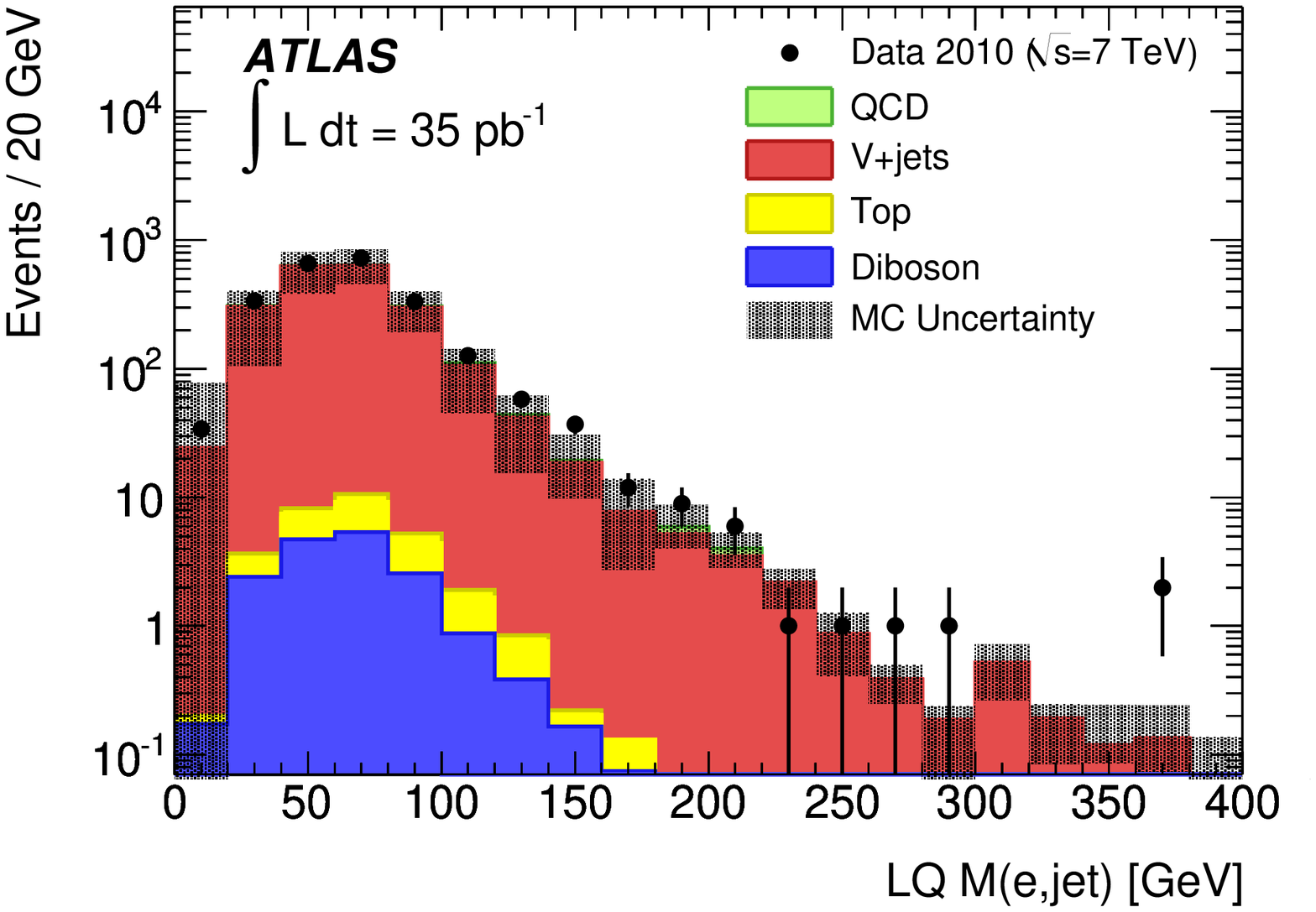}
   \includegraphics[width=3.5truein]{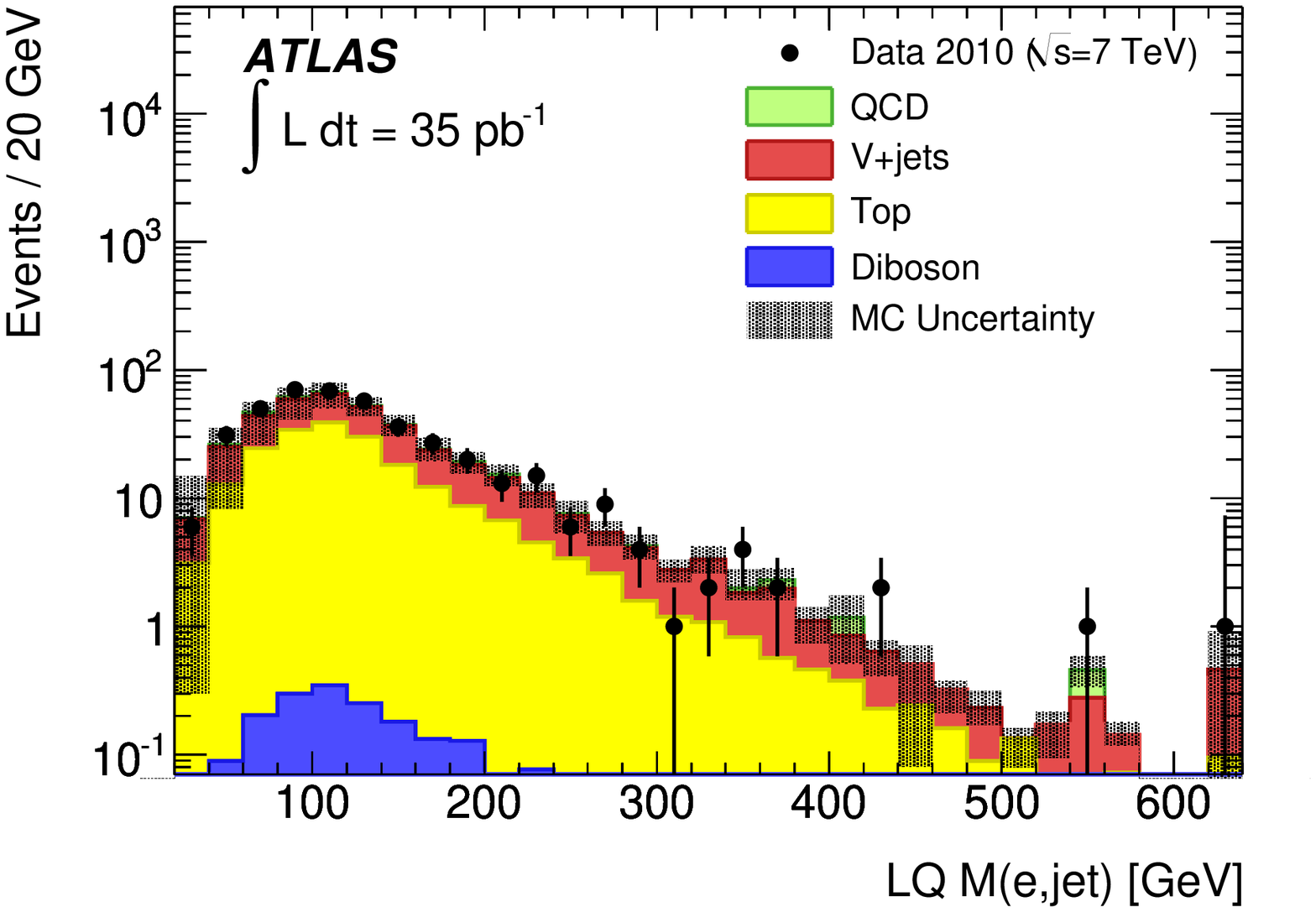}
   \includegraphics[width=3.5truein]{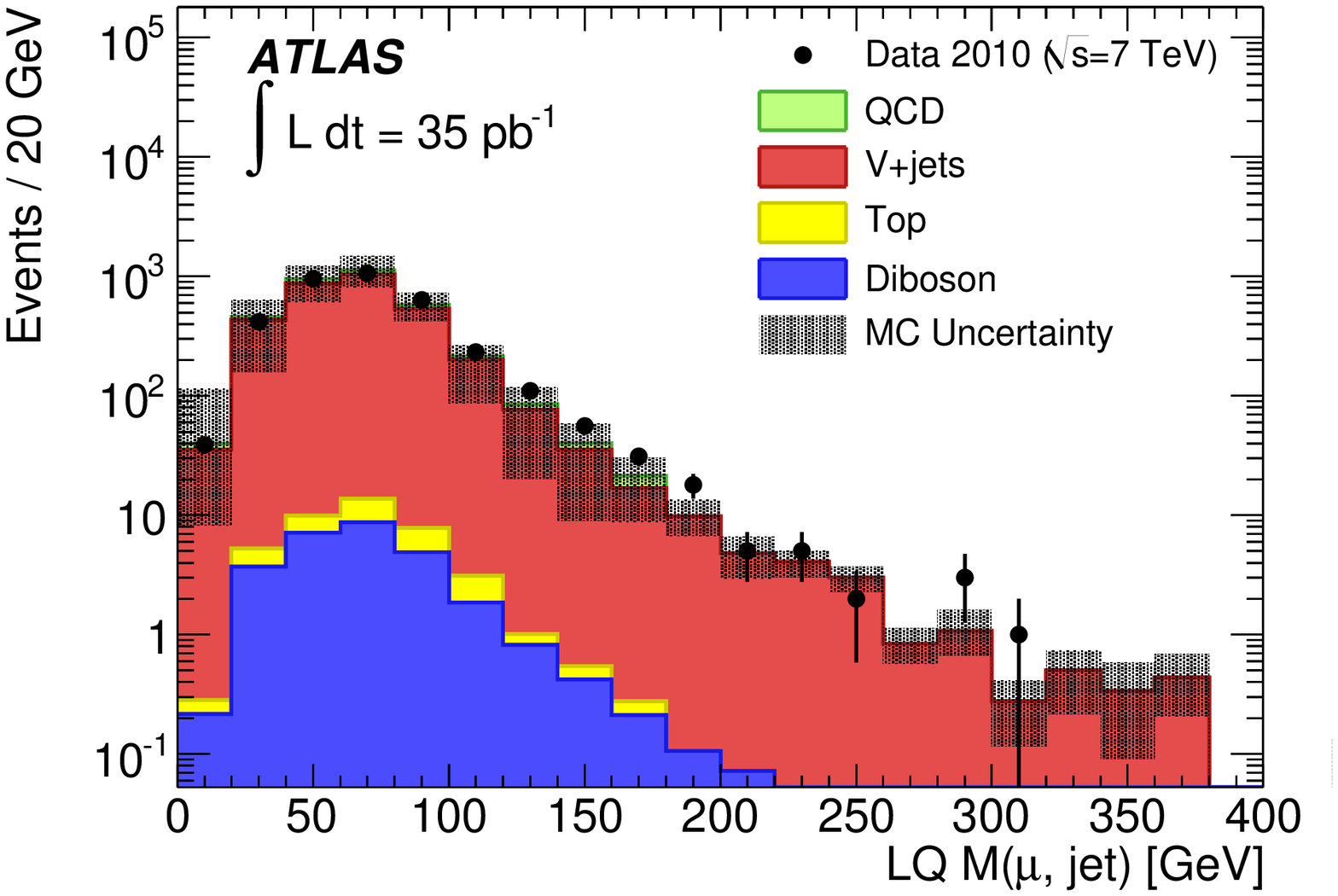}
  \includegraphics[width=3.5truein]{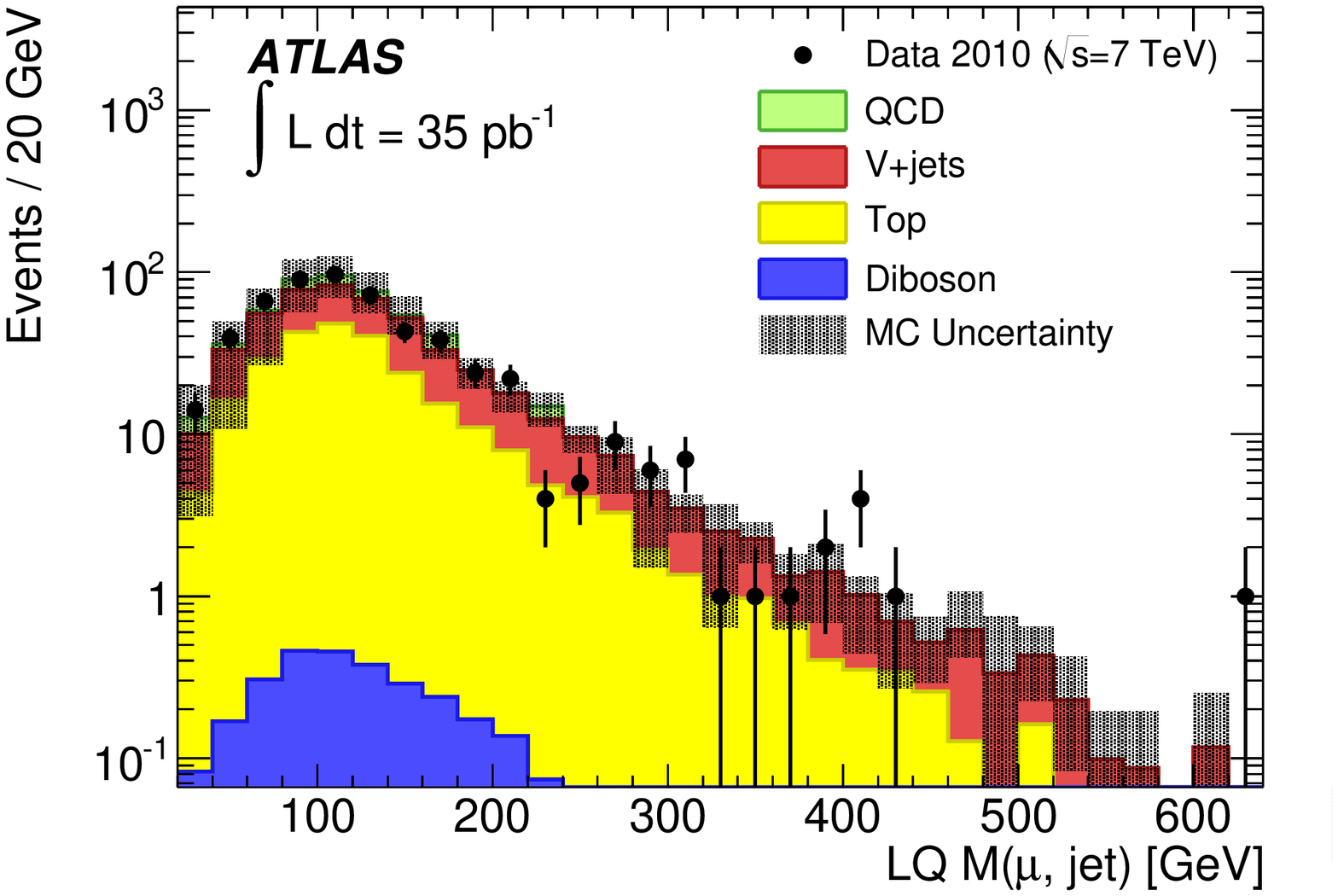}
  \caption{Reconstructed $\mlq$ distributions for the
   $\enjj$ (top) and the $\mnjj$ (bottom) final states in the $W+2$~jets (left) and $\ttbar$ (right) control regions.
The data are indicated by the points and the Standard Model backgrounds are shown with cumulative distributions.
The QCD background is estimated from data, while the other background contributions
are obtained from simulated samples. The top background includes both $\ttbar$ and single top events.
The expected contribution from a potential LQ signal is negligible
and is not included in the figure. The MC uncertainty band indicates the combined statistical 
and systematic uncertainties on the prediction in each bin.
   \label{f-encr}}
\end{figure*}

\section{Selection optimization}
The signal regions in the four final states are defined using a random grid search 
optimization method~\cite{Amos:1995tn}. The optimization is performed
using simulated signal and background samples (determined as described in Sections~VII and VIII) after preselection, with the final cuts
determined by minimizing the Poisson probability that the predicted background
fluctuates to at least the predicted signal plus background.  A
10\% overall scale uncertainty  on the yields, roughly matching the uncertainty on the luminosity,
is included in the optimization
as a systematic uncertainty. Optimization for the single lepton (dilepton) channels nominally is done using
signal samples with LQ masses of 300 (400)~GeV. Cuts derived using different LQ masses typically differ from the nominal values within statistical uncertainties. A variety of input variables are considered.
The requirements
that give the best separation between
signal and background are shown in Table~\ref{t-optvar}.
The yields obtained after final selection for the various channels are shown in 
Table~\ref{t-optsel-e}, along with the predicted background and signal.

\begin{table*}[hp]
\begin{center}
\caption{The selection requirements used to define the signal
region, as obtained from the optimization procedure. $\ptall >$ 30 GeV implies that the
$\pt$ of both the two leading leptons and the two leading jets in the dilepton samples should exceed 30 GeV.
 \label{t-optvar}}
\begin{tabular}{c|c|c}
	   $\eejj$ and  $\mmjj$  &  $\enjj$     & $\mnjj$  \\ \hline
   $\mll > 120$~GeV          & $\mt > 200$~GeV     & $\mt > 160$~GeV \\
   $\overline{\mlq}>150$~GeV & $\mlq>180$~GeV    & $\mlq>150$~GeV \\
   $\ptall>30$~GeV          & $\mtlq>180$~GeV  & $\mtlq>150$~GeV \\
   $\stl>450$~GeV               & $\stnu>410$~GeV       & $\stnu>400$~GeV \\
  \end{tabular}
\end{center}
\end{table*}

\begin{table*}
\begin{center}
  \caption{The predicted and observed yields in the signal region for all
    channels.  The $\lljj$ ($\lnjj$) channel signal yields are computed assuming
    $\beta=1.0\,\,(0.5)$. 
Both statistical and systematic uncertainties are included.\  Top refers to both single top and $\ttbar$ events.
    \label{t-optsel-e}}
  \begin{tabular}{l||rcl|rcl||rcl|rcl}
   Source  & \multicolumn{3}{c}{$eejj$} & \multicolumn{3}{c}{$\enjj$}& \multicolumn{3}{c}{$\mmjj$} & \multicolumn{3}{c}{$\mnjj$}  \\ \hline
   $V$+jets       & $0.50$&$\pm$&$0.28$ & $0.65$&$\pm$&$0.38$ & $0.28$&$\pm$&$0.22$ & $2.6 $&$\pm$&$1.4$\\
   Top            & $0.51$&$\pm$&$0.23$& $0.67 $&$\pm$&$0.39$ & $0.52$&$\pm$&$0.23$ & $1.6 $&$\pm$&$0.9$\\
   Diboson        & $0.03$&$\pm$&$0.01$ & $0.10 $&$\pm$&$0.03$ & $0.04$&$\pm$&$0.01$ & $0.10 $&$\pm$&$0.03$\\
   QCD            & $0.02$&$^{+}_{-}$&$^{0.03}_{0.02}$ &$0.06 $&$\pm$&$0.01$ & $0.00$&$^{+}_{-}$&$^{0.01}_{0.00}$ & $0.0 $&$\pm$&$ 0.0$\\ \hline
   {\bf Total Bkg}& $1.1$&$\pm$&$0.4$ & $1.4 $&$\pm$&$ 0.5$ &$0.8$&$\pm$&$0.3$& $4.4 $&$\pm$&$1.9$\\
   Data           & \multicolumn{3}{c|}{$2$}         & \multicolumn{3}{c||}{$2$}             & \multicolumn{3}{c|}{$0$}             & \multicolumn{3}{c}{$4$}\\ \hline
   LQ(250 GeV)    & $38$&$\pm$&$8$ & $9.6 $&$\pm$&$2.1$ &$45$&$\pm$&$10$   & $13 $&$\pm$&$3$\\
   LQ(300 GeV)    & $17$&$\pm$&$4$ & $5.1 $&$\pm$&$1.1$ & $21$&$\pm$&$5$  & $6.4 $&$\pm$&$1.4$\\
   LQ(350 GeV)    & $7.7$&$\pm$&$1.7$ & $2.6 $&$\pm$&$0.6$ & $9.4$&$\pm$&$2.1$   & $3.0 $&$\pm$&$0.7$\\
   LQ(400 GeV)    & $3.5$&$\pm$&$0.8$ & \multicolumn{3}{c||}{---}             & $4.4$&$\pm$&$1.0$  &   \multicolumn{3}{c}{---}                       \\
  \end{tabular}
\end{center}
\end{table*}

\section{Systematic uncertainties}
Systematic uncertainties are derived for a variety of sources, including data-simulation
differences in trigger and reconstruction efficiencies and 
 in the energy and momentum resolutions for leptons, jets and $\MET$,
instantaneous and integrated luminosity, modeling of the underlying event, 
variations in the method used to determine the QCD background,
and uncertainty on the LQ pair-production cross section.
A summary of the systematic uncertainties for the four final states is shown in
Table~\ref{t-ee-syssum}.

\begin{table*}[ht!]
\begin{center}{
  \caption{Systematic uncertainties for the $\lljj$ and the $\lnjj$ final
     states. The lepton trigger, identification and momentum (energy) scale and
    resolution uncertainties are small and grouped together. Single top and $\ttbar$ events are also grouped together. Uncertainties marked with $*$ ($**$) are 
only for the electron (muon) channels. QCD modeling systematics are not shown. All numbers are in percentages. ~\label{t-ee-syssum}}
     \footnotesize
  \begin{tabular}{l|cc|cc|cc|cc}
                          & \multicolumn{2}{c|}{$V$+jets} & \multicolumn{2}{c|}{Top} & \multicolumn{2}{c|}{Diboson} & \multicolumn{2}{c}{LQ (300~GeV)} \\
 Channel                  & $\lljj$     &  $\lnjj$       & $\lljj$      &  $\lnjj$    & $\lljj$      &  $\lnjj$     & $\lljj$      &  $\lnjj$ \\  \hline
 Production Cross Section & ---         & 4            & 13 & 13 &  5         & 5          & 18          & 18  \\
 Modeling                 & 34$^*$, 45$^{**}$        & 40           & 35         & 35         & ---          & ---          & ---          & --- \\
 Electron Energy Scale \& Resolution$^*$     & +13, -0.2 &$5$        & 10         & $2$      & 7          & 1          & 8          & $1$ \\
 Muon Momentum Scale \& Resolution$^{**}$& 20  &$5$       & 7          & $2$      & 8          & 1         & 6.7          & $1$ \\
 Jet Energy Scale         & 6         & +22, -13    & +9, -18  & 32 & +16, -6 & +17, -24   & 2          & $3$ \\
 Jet Energy Resolution    & 16        & 10           & 0.3        &  26       & 4          & 14          & 0.3        & $3$ \\
 Luminosity               & 0.3       & 11           & 11         & 11        & 11          & 11        & 11         & 11 \\
 Pile up                  & $<0.1$         & 5            & $<0.1$          & 4         &$<0.1$          & 6          & $<0.1$          & 2 \\ \hline
 Total Systematics        & 39$^*$   &+49, -45  & 47$^*$        & 57            & (+22, -16)       & +26,- 31 & 22              & 22 \\
                          & 52$^{**}$            &             &  (+49 -44)$^{**}$  &                 &    &  &               &  \\
   \end{tabular}}
\end{center}
\end{table*}

The lepton trigger and reconstruction efficiency systematic uncertainties are
derived by varying the selection of the $Z$ event sample used to
measure {\em in situ} efficiencies, and by varying the treatment of the (small)
background in this sample.  In addition, a 1\% uncertainty is included on the
muon isolation requirement that accounts for the difference of the efficiency of the isolation
requirement in data and $\ttbar$ and LQ simulated samples. Finally, a 3\% uncertainty 
is added to account for differences
in the muon $|d_0|$ distributions between data and simulation.  
Lepton momentum scale and resolution uncertainties are obtained by comparing the peak and width 
of $Z\to\ell\ell$ events in data and Monte Carlo.

The jet energy scale and resolution are varied by their uncertainties~\cite{JESUncertaintyConf_new,JERnote} for all simulated events, and their impact estimated independently. In
addition a 5\% uncertainty is added in quadrature to the jet energy scale
uncertainties to account for differences in response for quark and gluon jets.
These variations in scale and resolution are also propagated to the $\MET$.
The systematic uncertainty from instantaneous luminosity effects is evaluated
by comparing the results from simulated samples with and without additional minimum
bias events (pileup) added. This is an overestimate of the systematic,
but is still small (2-6\%, depending on the sample) compared 
to other sources.

Systematic uncertainties on the QCD background are determined by
comparing results from alternate normalizations to those 
described in Section~VII. 
The uncertainty on the $\lljj$ final states is determined
by varying the estimate between the nominal prediction and the upper level from the
extrapolation method described in Section~VII.  The uncertainty in the $\enjj$ final state is 
determined by comparing the default method, based on fits to the  $\mt$ distribution, to
alternate fits using the $\MET$ or the electron $\et$ distributions. The largest 
fractional difference (22\%) between
the nominal fit and fits using the alternate variables is taken as the systematic uncertainty. 
The uncertainty in the background to the $\mnjj$ final state is determined by 
comparing the default method to one that uses the muon isolation instead of the
$d_0$ variable
The difference in yield between the two estimates is used to 
determine the systematic uncertainty, giving 27\%.

The systematic uncertainties for the production models of $W$+jets and $Z$+jets
events in the single lepton analysis are determined by varying parameters at the
generator--level as described in~\cite{b-xsec-Wj-pub}, including
variations of the renormalization and factorization scales and minimum parton $\pt$.
The largest absolute deviation (40\%) is used as a
systematic uncertainty. In the dilepton analysis, the systematic uncertainty is determined by 
varying the
 dilepton mass window and replacing the ALPGEN event generator with SHERPA.
The difference between ALPGEN and SHERPA dominates, giving an 
uncertainty in the $\eejj(\mmjj)$ channel of 34 (45)\% on the
 $Z$+jets prediction in the signal region.

Systematic uncertainties are evaluated for the $\ttbar$ production model 
by comparing the predictions obtained with
MC@NLO and POWHEG. 
The result is a 35\% uncertainty for both $\lljj$ and $\lnjj$ channels.

An integrated luminosity uncertainty of 11\%~\cite{lum} is applied to all backgrounds
determined from simulated events and to the signal. Additional signal
uncertainties are obtained from varying the renormalization
scale parameter to 0.5$\mlq$ and to 2$\mlq$ (15\%),
from different PDF choices (13-17\%) for LQ masses in the range 300-400~GeV, 
and from Initial- and Final-State-Radiation effects (2\%). The systematic uncertainties on the
theory cross sections used to normalize backgrounds are given in Section IV.

\section{Results}
The expected and observed numbers of events are shown in Table~\ref{t-optsel-e}.
Data and Standard Model expectations are in good agreement.
95\% CL upper limits on LQ pair--production cross sections are determined
using a modified frequentist approach~\cite{b-lcalc1,b-lcalc2}. 
Systematic uncertainties are incorporated in the limit calculation
as nuisance parameters and are integrated out.
The limits are translated into bounds in the $\beta$ versus LQ mass plane.

Monte Carlo studies show
that a better sensitivity is achieved when using 
kinematic shapes rather than just the total number of events.
For the $\eejj$ and $\mmjj$ final states, the observed and predicted $\st$
distributions are used in the limit-setting procedure, and for the $\enjj$ and $\mnjj$
final states the observed and predicted $\mlq$ distributions are
used. The $\st$ distributions are shown in Fig.~\ref{f-eesig} for the
$\lljj$ channels. Figure~\ref{f-ensig} shows the $\mlq$ distributions 
for the single lepton final states. 

\begin{figure*}[h!]
\includegraphics[width=3.5truein]{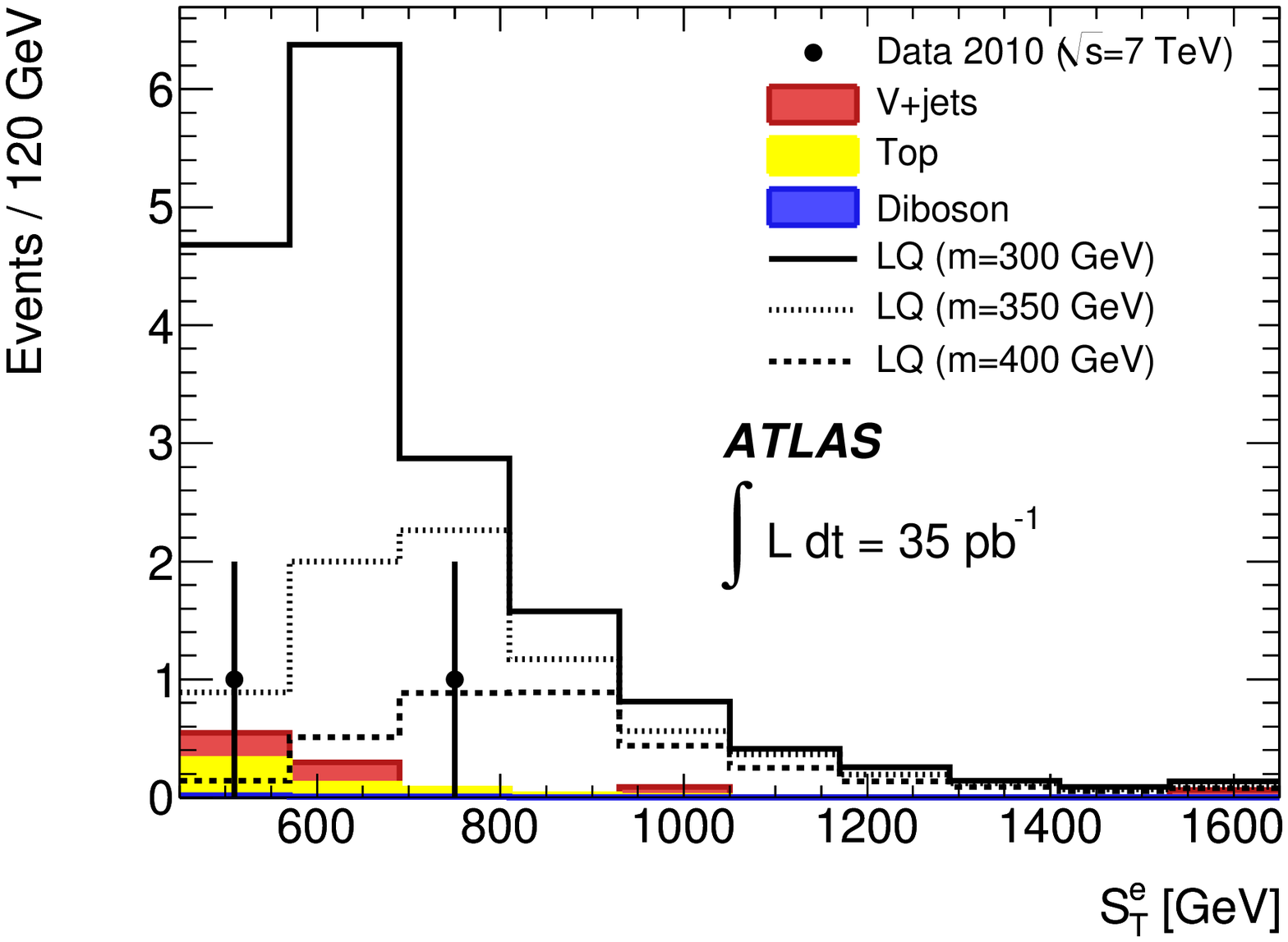}
\includegraphics[width=3.5truein]{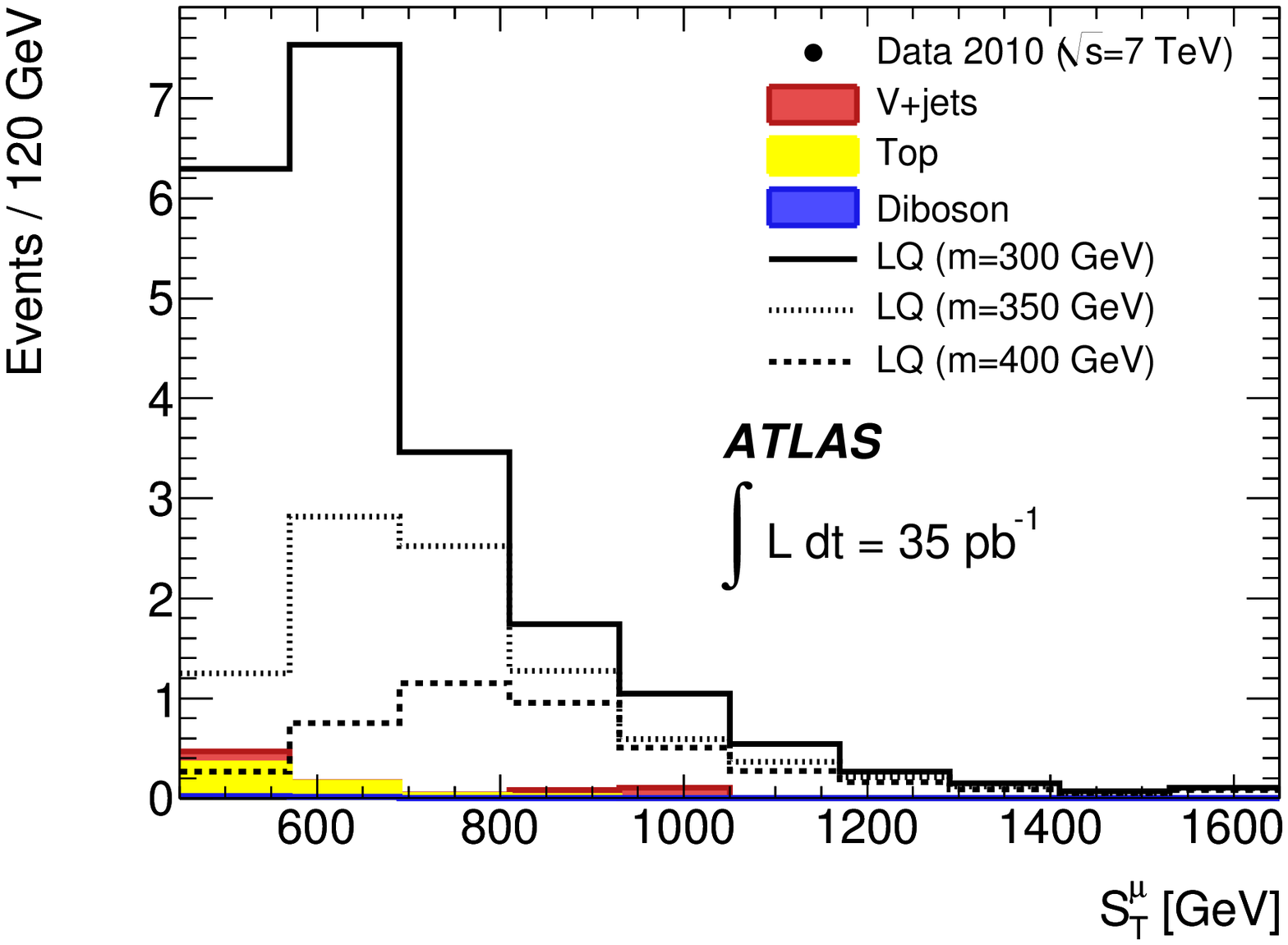}
 \caption{$\stl$ distribution for the $eejj$ (left) and the $\mmjj$ final states (right) after all selections. The data are indicated by the points and the Standard Model backgrounds are shown with cumulative distributions. The expected LQ signals for various masses are also shown.\label{f-eesig}}
\end{figure*}
\begin{figure*}[h!]
\includegraphics[width=3.5truein]{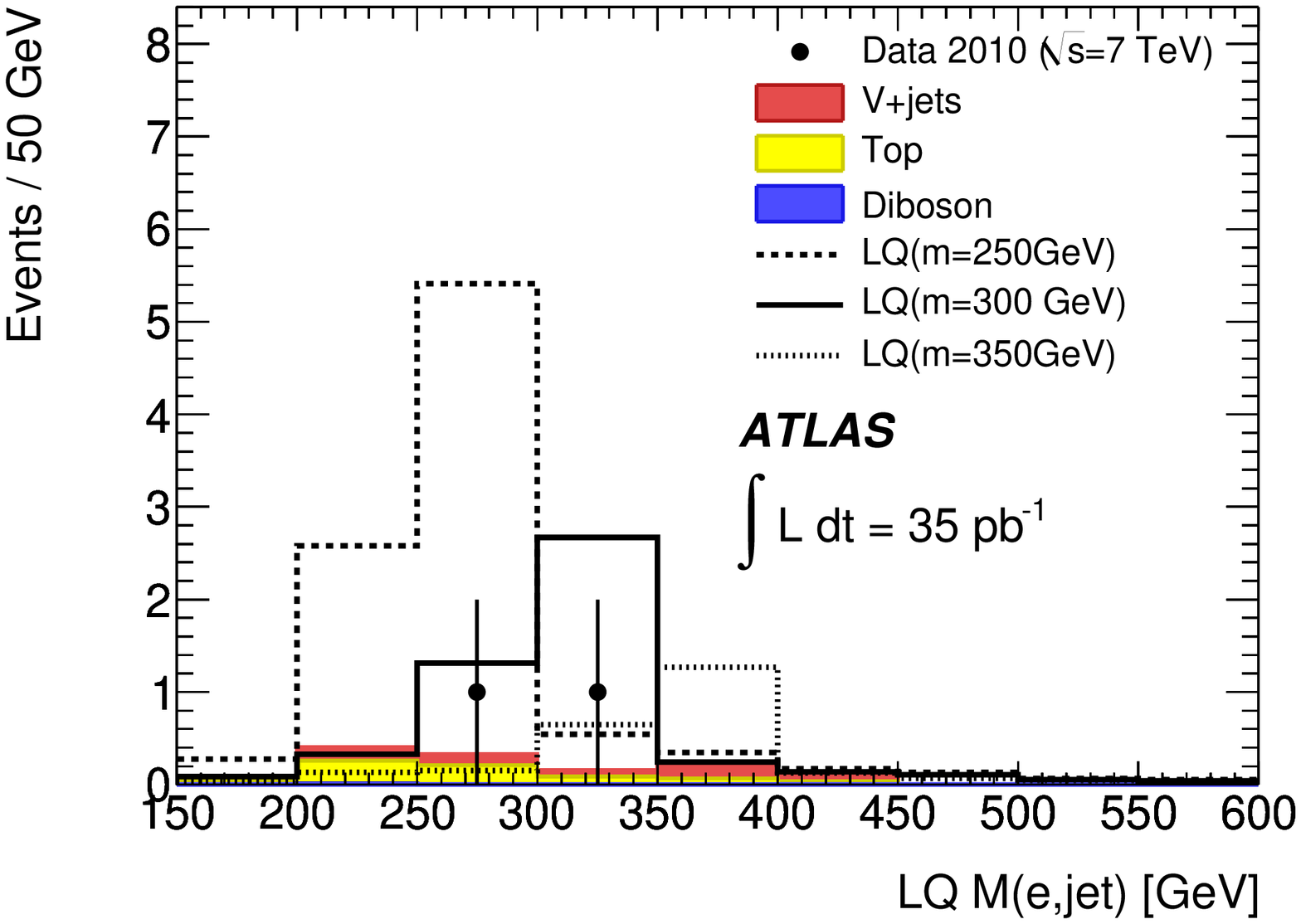}
\includegraphics[width=3.5truein]{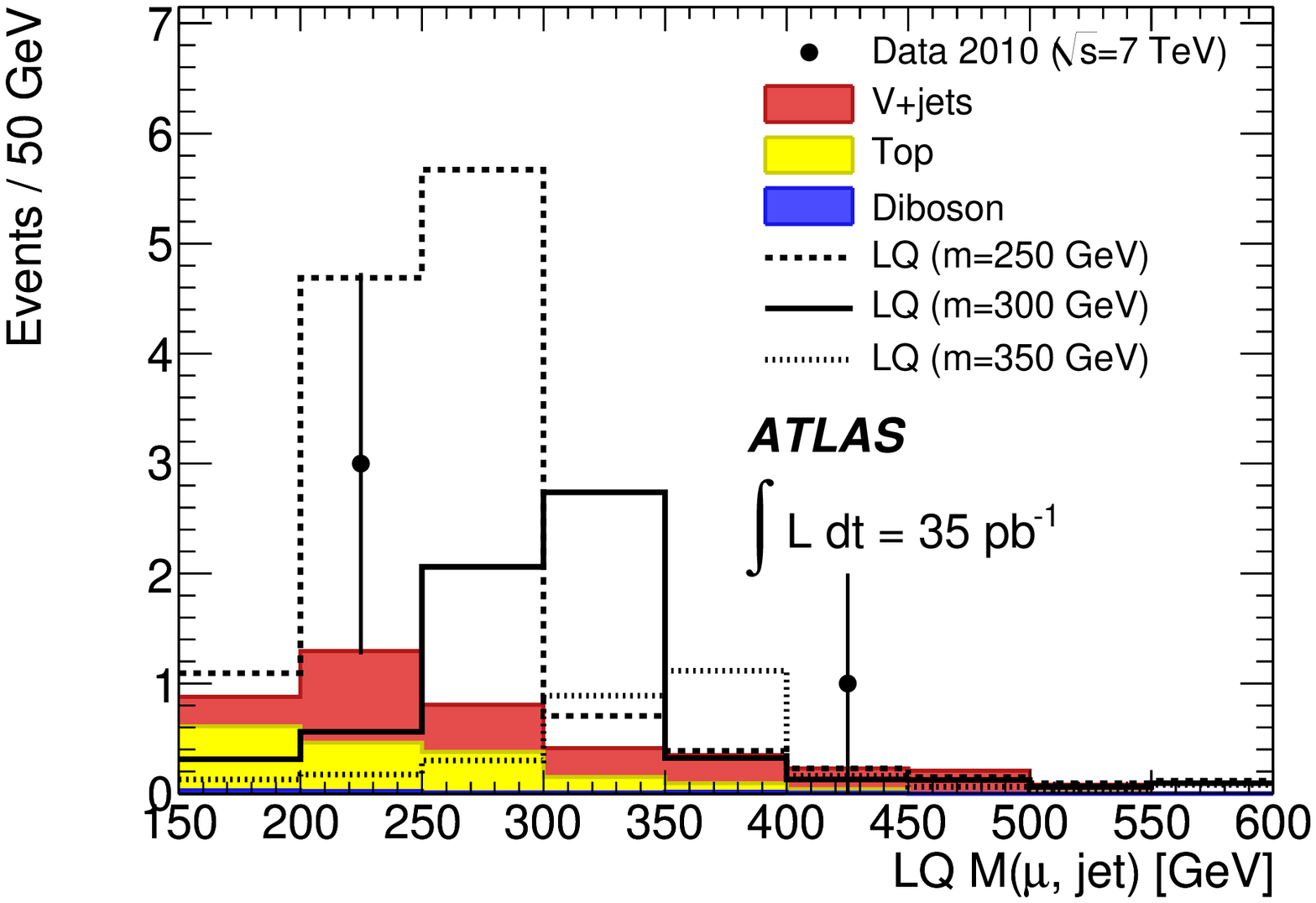}
  \caption{$\mlq$ distribution for the $\enjj$ (left) and the $\mnjj$ final state (right) after all selections. The data are indicated by the points and the Standard Model backgrounds are shown with cumulative distributions. The expected LQ signals for various masses are also shown.\label{f-ensig}}
\end{figure*}

The 95\% CL upper bounds on the cross section for first and second generation LQ pair
production as a function of mass are shown in Fig.~\ref{f-xsec95-e} for $\beta=$1.0 and
$\beta=$0.5. The combined limits are also shown in the $\beta$
vs. $\mlq$ plane in Fig.~\ref{f-bvm-e} for both generations.
The expected and observed combined limits including all systematic uncertainties 
are shown in Table~\ref{t-limitscombined} for both $\beta=$1.0 and 0.5. The systematic
uncertainties, of the order of 50\%, only change the limits by 5 to 10~GeV, depending
on the value of $\beta$.

\begin{figure*}[h]
\includegraphics[width=3.5truein]{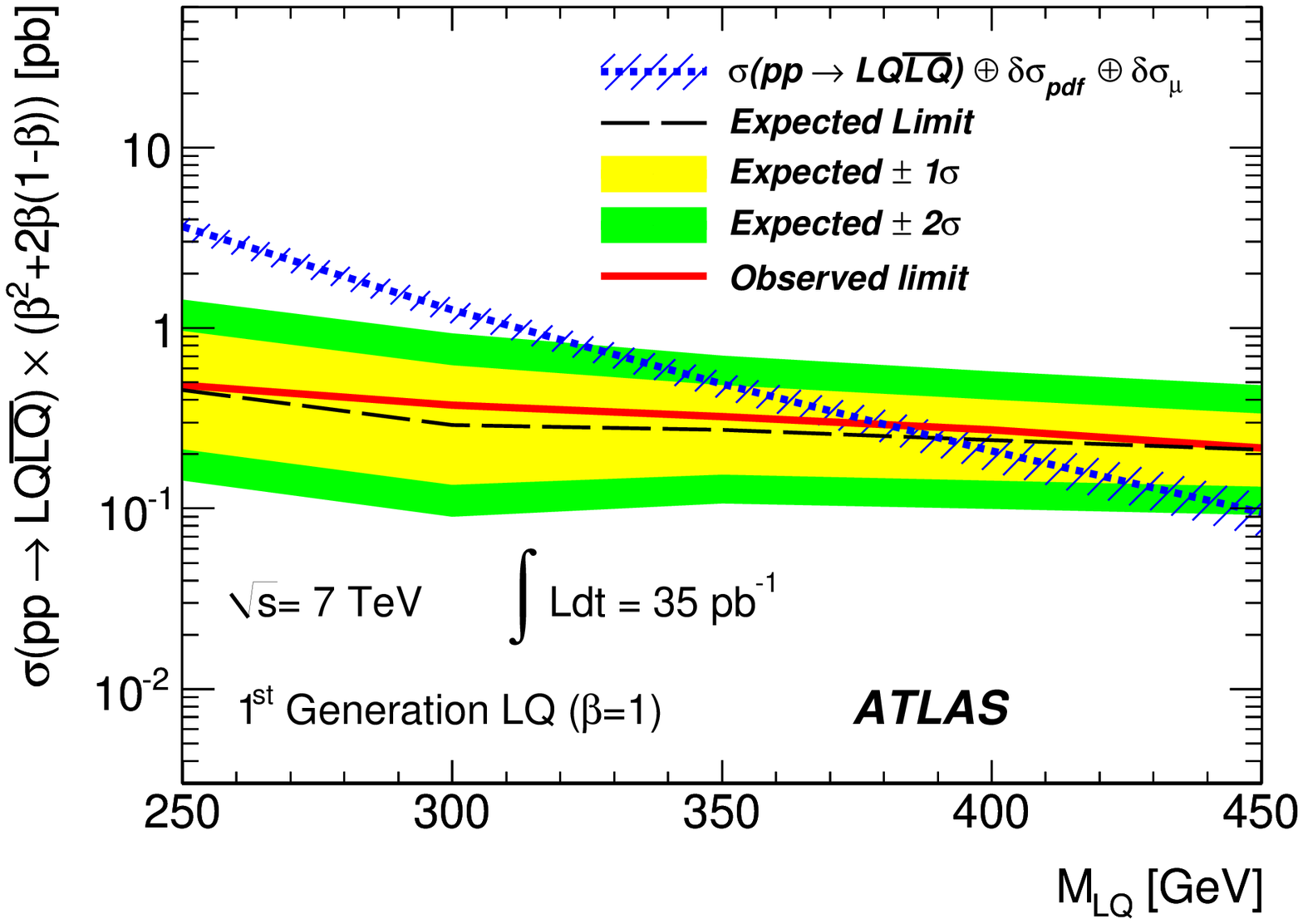}
\includegraphics[width=3.5truein]{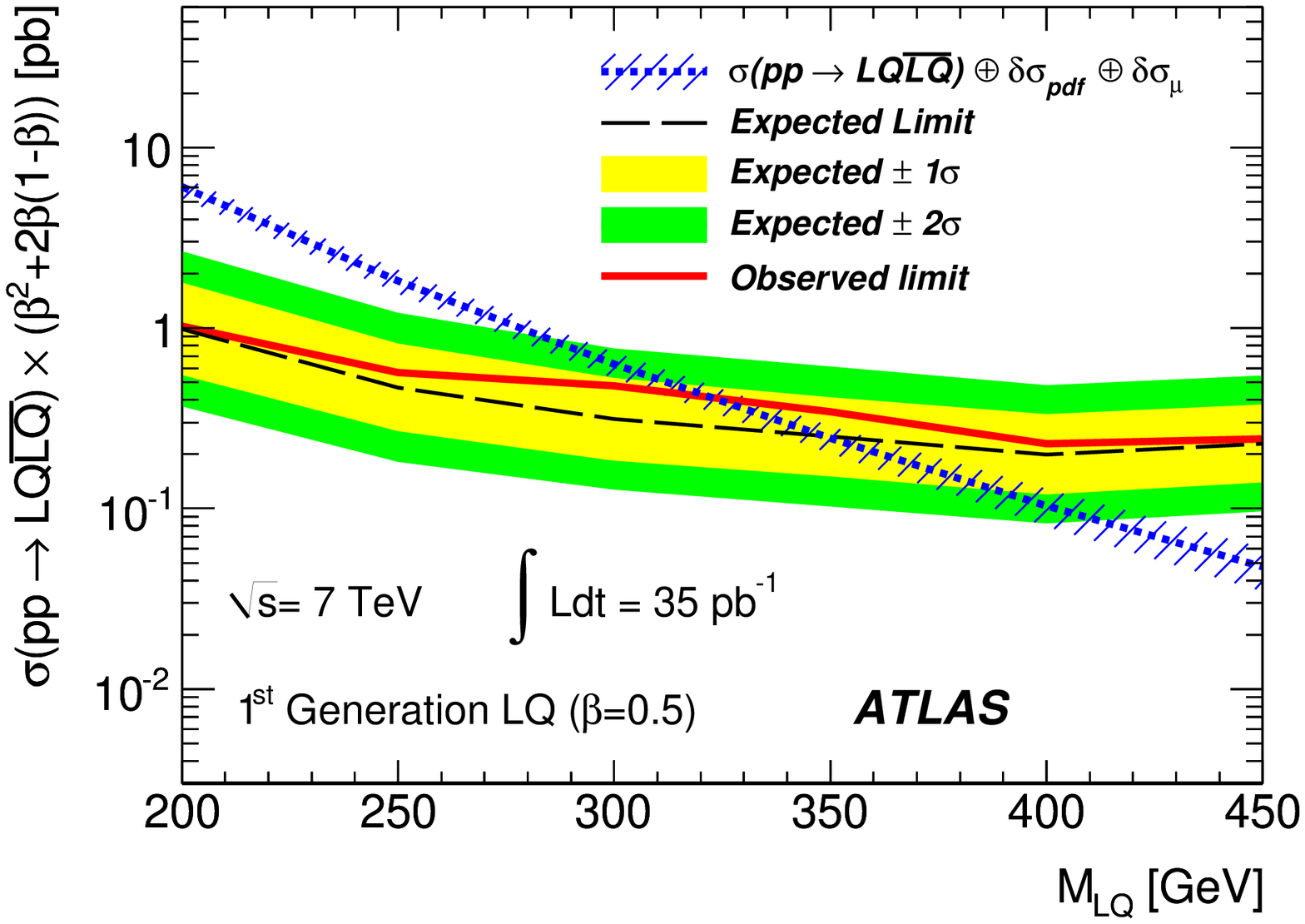}
\includegraphics[width=3.5truein]{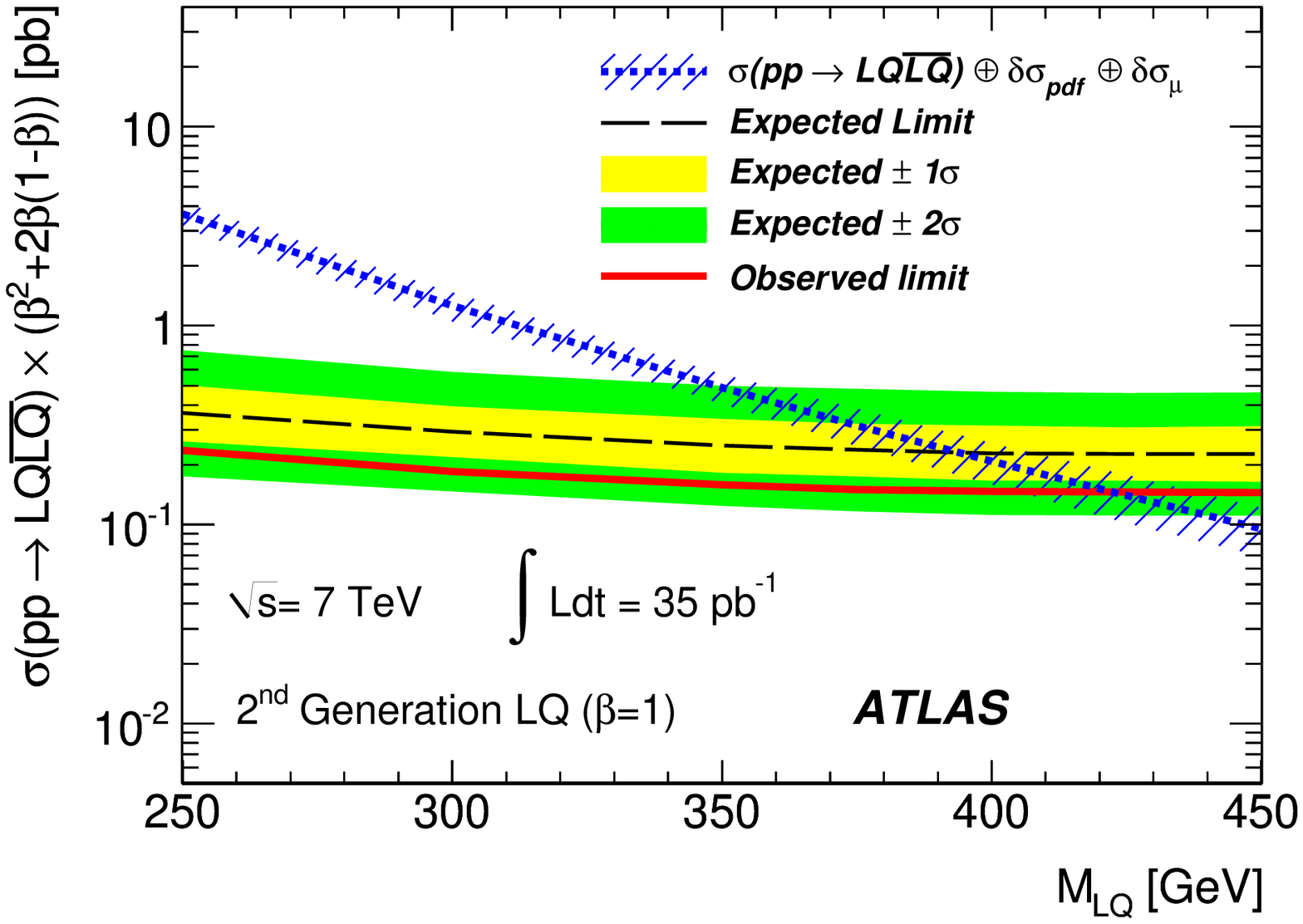}
 \includegraphics[width=3.5truein]{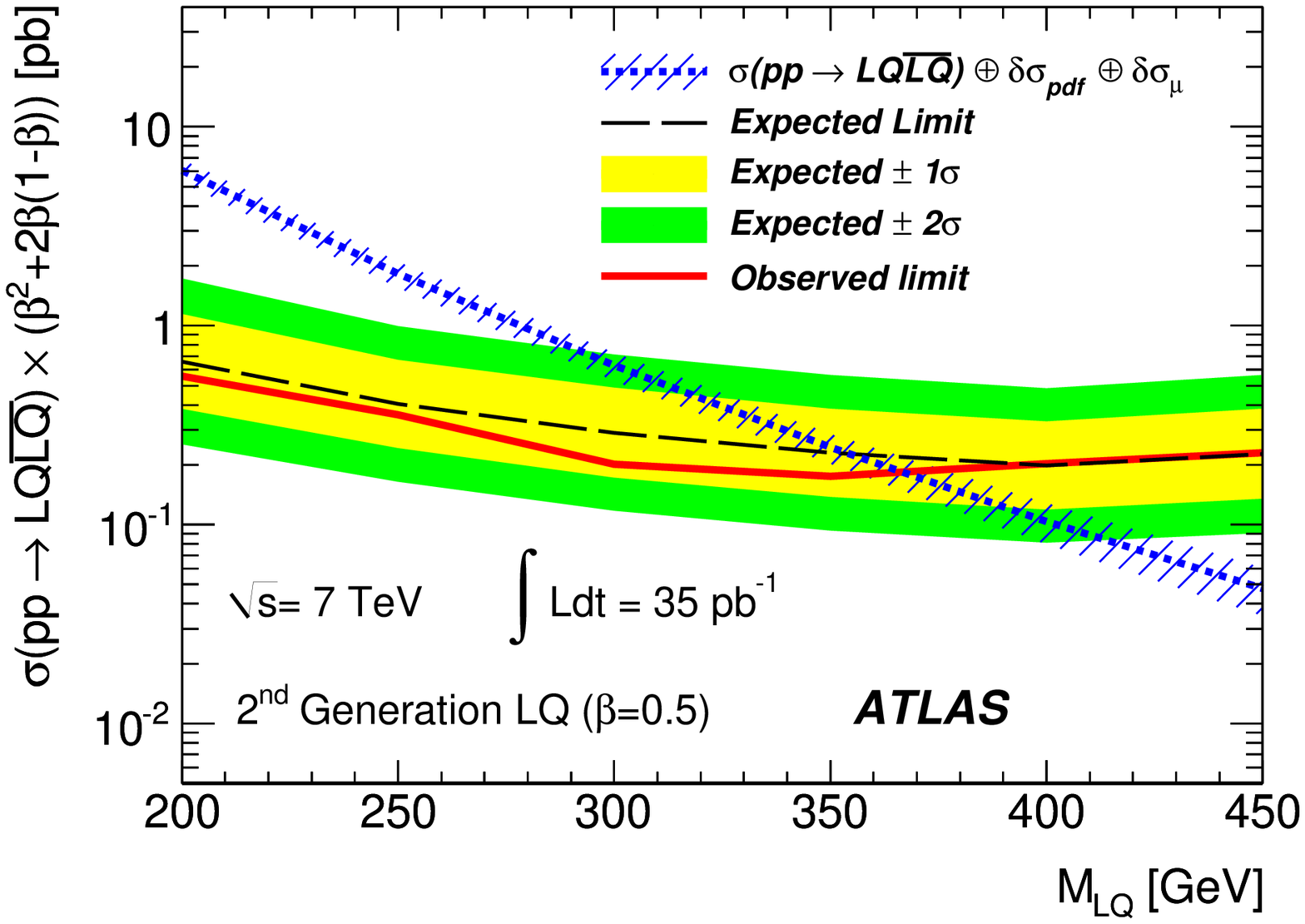}
 \caption{95\% CL upper bounds on the LQ pair production cross sections as a
  function of LQ mass for the combination of the $\eejj$ and the $\enjj$ channels (top),
and for the $\mmjj$ and the $\mnjj$ channels (bottom).
   The figures on the left (right) show the limit for $\beta=$1.0 (0.5). The expected limit is indicated by the dashed line. 
The light yellow (dark green) solid band contains 68\% (95\%) of possible outcomes from pseudo-experiments in which
the yield is Poisson-fluctuated around the background-only expectation. Systematics uncertainties are included.
 The observed limit
   is indicated by the solid line.
The theory prediction
   is indicated by the dotted line, which includes systematic uncertainties due to the choice
   of the PDF and to the renormalization and factorization scales ($\Delta\sigma_\mu$).\label{f-xsec95-e}}
\end{figure*}
\begin{figure*}[h]
\includegraphics[width=3.5truein]{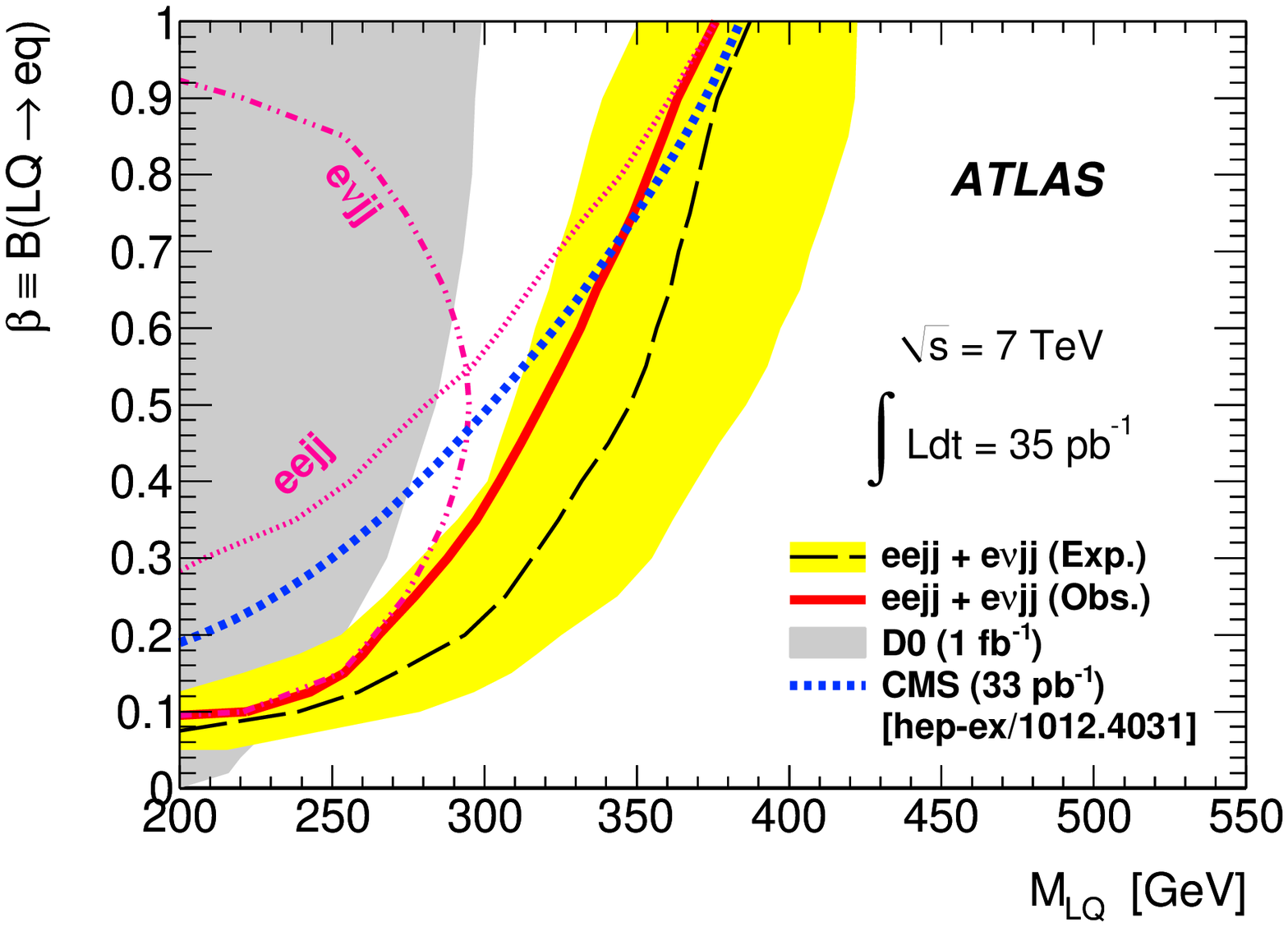}
  \includegraphics[width=3.5truein]{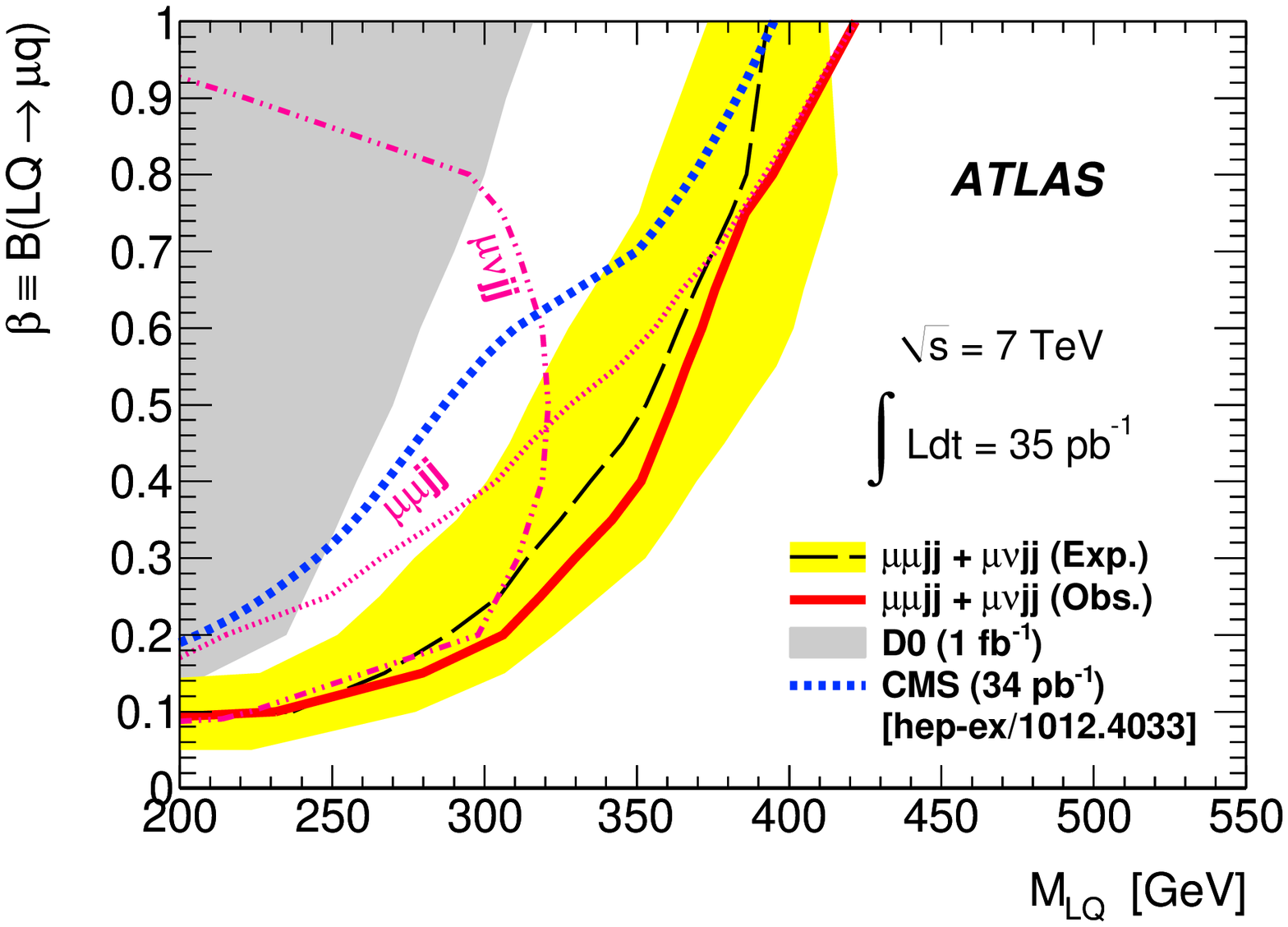}
  \caption{95\% CL exclusion region obtained from the combination of the
  two electron channels (left) and the muon channels (right)  
shown in the $\beta$ versus leptoquark mass plane. The gray
  area indicates the D0 exclusion limit~\cite{results_first_gen_limit_d0,results_second_gen_limit_d0}, and the thick dotted line the CMS exclusion~\cite{results_first_gen_limit_cms,results_second_gen_limit_cms}. 
The dotted and dotted-dashed lines
  show the individual limits for the $\lljj$ and the $\lnjj$ channels, respectively.
  The combined expected limit is indicated by the thick dashed line. 
The solid band contains 68\% of possible outcomes from pseudo-experiments in which
the yield is Poisson-fluctuated around the background-only expectation. Systematic uncertainties are included.
The combined observed
  limit is indicated by the solid line.\label{f-bvm-e}}
\end{figure*}

\begin{table*}[hp]
\begin{center}
\caption{95\% CL expected and observed limits on the first and second generation LQs under different
assumptions on $\beta$. The expected limits include systematic uncertainties.\label{t-limitscombined}}
\begin{tabular}{c|c|c}
Type ($\beta)$ & Expected limit (GeV) & Observed limit (GeV)\\
\hline
1st generation (1.0) & 387 & 376 \\
1st generation (0.5) & 348 & 319 \\
2nd generation (1.0) & 393 & 422 \\
2nd generation (0.5) & 353 & 362 \\
\end{tabular} 
\end{center}
\end{table*}

\section{Conclusions}
This paper reports the results of searches for pair production of first or second
generation scalar leptoquarks using a data sample corresponding to an
integrated luminosity of 35~pb$^{-1}$. 
The data in the high
signal--to--background signal region are in good agreement with the
Standard Model expectations. 95\% CL upper bounds on the production cross section are
determined.  These are translated into lower bounds for first (second) generation leptoquark masses of
$\mlq>376$ (422)$~$GeV and $\mlq>319$ (362)$~$GeV for $\beta=1.0$ and $\beta=0.5$, respectively. These
are the most stringent bounds to date from direct searches for leptoquarks in much of the phase space.

\section{Acknowledgements}

We thank CERN for the very successful operation of the LHC, as well as the
support staff from our institutions without whom ATLAS could not be
operated efficiently.

We acknowledge the support of ANPCyT, Argentina; YerPhI, Armenia; ARC,
Australia; BMWF, Austria; ANAS, Azerbaijan; SSTC, Belarus; CNPq and FAPESP,
Brazil; NSERC, NRC and CFI, Canada; CERN; CONICYT, Chile; CAS, MOST and
NSFC, China; COLCIENCIAS, Colombia; MSMT CR, MPO CR and VSC CR, Czech
Republic; DNRF, DNSRC and Lundbeck Foundation, Denmark; ARTEMIS, European
Union; IN2P3-CNRS, CEA-DSM/IRFU, France; GNAS, Georgia; BMBF, DFG, HGF, MPG
and AvH Foundation, Germany; GSRT, Greece; ISF, MINERVA, GIF, DIP and
Benoziyo Center, Israel; INFN, Italy; MEXT and JSPS, Japan; CNRST, Morocco;
FOM and NWO, Netherlands; RCN, Norway; MNiSW, Poland; GRICES and FCT,
Portugal; MERYS (MECTS), Romania; MES of Russia and ROSATOM, Russian
Federation; JINR; MSTD, Serbia; MSSR, Slovakia; ARRS and MVZT, Slovenia;
DST/NRF, South Africa; MICINN, Spain; SRC and Wallenberg Foundation,
Sweden; SER, SNSF and Cantons of Bern and Geneva, Switzerland; NSC, Taiwan;
TAEK, Turkey; STFC, the Royal Society and Leverhulme Trust, United Kingdom;
DOE and NSF, United States of America.

The crucial computing support from all WLCG partners is acknowledged
gratefully, in particular from CERN and the ATLAS Tier-1 facilities at
TRIUMF (Canada), NDGF (Denmark, Norway, Sweden), CC-IN2P3 (France),
KIT/GridKA (Germany), INFN-CNAF (Italy), NL-T1 (Netherlands), PIC (Spain),
ASGC (Taiwan), RAL (UK) and BNL (USA) and in the Tier-2 facilities
worldwide.

\bibliography{lqlq}

\begin{thebibliography}{52}
\bibitem[{foo({\natexlab{a}})}]{footnote1}
\bibinfo{howpublished}{{The $LQ-\ell-q$ coupling determines the LQ lifetime and
  width. For LQ masses in the range considered here, $200$~GeV$\le \mlq \le
  400$~GeV, couplings greater than $q_e\times 10^{-6}$, with $q_e$ the electron
  charge, correspond to decay lengths less than roughly 1~mm. In addition, to
  be insensitive to the coupling, the width cannot be larger than the
  experimental resolution, of a few GeV. This sets the approximate sensitivity
  to the unknown coupling strength.}}
\bibitem[{foo({\natexlab{b}})}]{footnote3}
\bibinfo{howpublished}{{By convention we place the electron and its neutrino,
  and the up and down quarks in the first generation.}}
\bibitem[{foo({\natexlab{c}})}]{footnote2}
\bibinfo{howpublished}{{ATLAS uses a right-handed coordinate system with its
  origin at the nominal interaction point (IP) in the centre of the detector
  and the $z$-axis coinciding with the axis of the beam pipe. The $x$-axis
  points from the IP to the centre of the LHC ring, and the $y$-axis points
  upward. Cylindrical coordinates ($r$,$\phi$) are used in the transverse
  plane, $\phi$ being the azimuthal angle around the beam pipe. The
  pseudorapidity is defined in terms of the polar angle $\theta$ as $\eta
  \equiv $-ln(tan($\theta$/2))}}.
\bibitem[{foo({\natexlab{d}})}]{footnote4}
\bibinfo{howpublished}{{For single lepton final states, events with fake
  leptons are purely multijet events. For dileptons, events with fake leptons
  come in two forms -- multijet events with two fake leptons, and single top
  and $W$+jets events with one real lepton and one fake lepton.}}
\end{thebibliography}

\clearpage
\begin{flushleft}
{\Large The ATLAS Collaboration}

\bigskip

G.~Aad$^{\rm 48}$,
B.~Abbott$^{\rm 111}$,
J.~Abdallah$^{\rm 11}$,
A.A.~Abdelalim$^{\rm 49}$,
A.~Abdesselam$^{\rm 118}$,
O.~Abdinov$^{\rm 10}$,
B.~Abi$^{\rm 112}$,
M.~Abolins$^{\rm 88}$,
H.~Abramowicz$^{\rm 153}$,
H.~Abreu$^{\rm 115}$
% ...
\bigskip

% ...
$^{10}$ Institute of Physics, Azerbaijan Academy of Sciences, Baku, Azerbaijan\\
$^{11}$ Institut de F\'isica d'Altes Energies and Universitat Aut\`onoma  de Barcelona and ICREA, Barcelona, Spain\\
% ...
$^{48}$ Fakult\"{a}t f\"{u}r Mathematik und Physik, Albert-Ludwigs-Universit\"{a}t, Freiburg i.Br., Germany\\
$^{49}$ Section de Physique, Universit\'e de Gen\`eve, Geneva, Switzerland\\
% ...
$^{88}$ Department of Physics and Astronomy, Michigan State University, East Lansing MI, United States of America\\
% ...
$^{111}$ Homer L. Dodge Department of Physics and Astronomy, University of Oklahoma, Norman OK, United States of America\\
$^{112}$ Department of Physics, Oklahoma State University, Stillwater OK, United States of America\\
% ...
$^{115}$ LAL, Univ. Paris-Sud and CNRS/IN2P3, Orsay, France\\
% ...
$^{118}$ Department of Physics, Oxford University, Oxford, United Kingdom\\
% ...
$^{153}$ Raymond and Beverly Sackler School of Physics and Astronomy, Tel Aviv University, Tel Aviv, Israel\\
% ...
\end{flushleft}

\end{document}